\documentclass{pasj00}
\usepackage{color}

\begin{document}
\SetRunningHead{Y. Takeda et al.}{O~{\sc i} 7771--5 triplet and microturbulence 
in evolved A-, F-, and G-stars}
\Received{2017/10/23}
\Accepted{2017/12/13}

\title{Luminosity effect of O~I 7771--5 triplet and atmospheric microturbulence 
in evolved A-, F-, and G-type stars
}

\author{
Yoichi \textsc{Takeda},\altaffilmark{1,2}
Gwanghui \textsc{Jeong},\altaffilmark{3,4}
and
Inwoo \textsc{Han},\altaffilmark{3,4}
}
\altaffiltext{1}{National Astronomical Observatory, 2-21-1 Osawa, Mitaka, Tokyo 181-8588, Japan}
\altaffiltext{2}{SOKENDAI, The Graduate University for Advanced Studies, 
2-21-1 Osawa, Mitaka, Tokyo 181-8588}
\email{takeda.yoichi@nao.ac.jp}
\altaffiltext{3}{Korea Astronomy and Space Science Institute,
776, Daedeokdae-Ro, Youseong-Gu, Daejeon 34055, Korea}
\altaffiltext{4}{Korea University of Science and Technology, 217, 
Gajeong-ro Yuseong-gu, Daejeon 34113, Korea}
\email{tlotv@kasi.re.kr, iwhan@kasi.re.kr}



\maketitle

\begin{abstract}
It is known that the strength of neutral oxygen triplet lines at 
7771--5~\AA\ shows a luminosity effect in evolved A through G stars.
However, its general behavior across the HR diagram is not yet well understood, 
since the applicability limit of the relations proposed by various previous 
work (tending to be biased toward supergiants) still remains unclear. 
Besides, our understanding on the nature of atmospheric micro-scale turbulence, 
which is considered to play a significant role (along with the non-LTE line 
intensification) for the cause of this effect, is still insufficient. 
Towards clarifying these problems, we carried out an extensive non-LTE 
spectrum-fitting analysis of O~{\sc i} 7771--5 lines for unbiased sample of 
75 evolved A-, F,- and G-type stars over wide luminosity classes (from subgiants 
through supergiants) including rapid rotators, from which the total equivalent 
width ($W_{77}$) was derived and the microturbulence ($\xi$) was determined 
by two different (profile- and abundance-based) methods for each star.
While we confirmed that $W_{77}$ tends to increase in the global sense 
as a star's absolute magnitude ($M_{V}$) becomes more luminous, distinctly
different trends were found between lower-gravity ($\log g \ltsim 2.5$) and 
higher-gravity ($\log g \gtsim 2.5$) stars, in the sense that the $M_{V}$ vs. $W_{77}$ 
formulas proposed by past studies are applicable only to the former supergiant group.
In case of using $W_{77}$ for empirical $M_{V}$ evaluation by such simple formulas, 
it is recommended to confine only to supergiants of $-5 \gtsim M_{V} \gtsim -10$. 
Regarding the microturbulence significantly controlling $W_{77}$, it roughly 
shows an increasing tendency with a decrease in surface gravity. However, the trend 
is not monotonic but rather intricate (e.g., hump, stagnation, or discontinuously 
large increase) depending on the stellar type and evolutionary stage.  
\end{abstract}

%


\section{Introduction}

Ever since the historical work of old days (Merrill 1925, Keenan \& Hynek 1950),
the triplet lines of neutral oxygen at 7771.94, 7774.17, and 7775.4 \AA\ 
(3s~$^{5}$S$^{\rm o}$--3p~$^{5}$P, multiplet 1) are known 
to be of considerable strength and easily measurable in the spectra of A-, F-, and G-type stars.
In particular, an important aspect related to this feature is the luminosity effect;
i.e., its strength (herein referred to as $W_{77}$, which is the total integrated 
equivalent width of the whole triplet) tends to progressively grow as a star becomes 
more luminous. It is natural to come up with a possibility to make use of this characteristic
to estimate the stellar absolute magnitude ($M_{V}$) by simply measuring $W_{77}$.
Accordingly, these triplet lines inspired the interest of a number of astrophysicists,  
who investigated the nature of $M_{V}$ vs. $W_{77}$ relation and established
useful analytical formulas, as summarized in table~1.\footnote{The references listed
in table~1 were selected by consulting the literature in the recent papers of 
Kovtyukh, Gorlova, and Belik (2012) as well as Dambis (2013). They are not meant to be complete.}

Unfortunately, our understanding of these $M_{V}$ vs. $W_{77}$ relations 
proposed by various investigators is not yet sufficient, since it appears 
still uncertain to which luminosity ranges such simple relations are applicable.\\
--- For example, Arellano Ferro, Giridhar, and Rojo Arellano (2003; the latest of the 
series of papers published by their group) reported that $M_{V}$ can be well represented 
by a quadratic polynomial of $W_{77}$ (with remarkable accuracies of only 
$\pm 0.38$~mag) for evolved A-, F, and G-stars over a wide luminosity range (from 
$M_{V} \sim $0~mag to $-9$~mag; cf. their Fig.~5b), in which even the color term 
[$(b-y)_{0}$] has been omitted (which was included in their earlier papers but 
concluded to be unnecessary). If this formula is valid, it must be very useful.\\
--- However, according to the recent investigation by Kovtyukh, Gorlova, and Belik (2012),
$M_{V}$ and $W_{77}$ do not seem to follow a simple and smooth one-to-one correspondence 
but show a considerable scatter, especially the lower luminosity region of 
$0 \gtsim M_{V} \gtsim -3$ (cf. their Fig.~2). Although they tried reproduce 
their observational $M_{V}$ vs. $W_{77}$ data by an analytical relation including 
terms involving $T_{\rm eff}$ (effective temperature), $\log g$ (surface gravity), 
$\xi$ (microturbulence), and [Fe/H] (metallicity), such a complex formula is not
practically useful as long as empirical determination of $M_{V}$ is concerned.

Looking over the various literature on the luminosity effect of O~{\sc i} 7771--5 (table~1),
we note that, while intrinsically bright ``supergiants'' were preferably investigated in almost 
all work, less attention has been paid to ``giants'' which are comparatively less luminous.
Accordingly, it seems necessary to have a better understanding of the behavior of
O~{\sc i} 7771--5 strength for A--G giants as well as supergiants, before we can answer 
the question ``Is $M_{V}$ simply a function of $W_{77}$ as reported by 
Arellano Ferro et al. (2003)?'' or ``Does it actually depend not only $W_{77}$ 
but also on other stellar parameters?''

Motivated by this situation, we decided to carry out a comprehensive study on the strength 
of O~{\sc i} 7771--5 lines for evolved A-, F-, and G-type stars of various luminosity 
classes (subgiants, giants, and supergiants) based on the high-dispersion spectra 
of an unbiased sample of 75 stars observed at Bohyunsan Astronomical Observatory, 
in order to clarify the parameter dependence of $W_{77}$ across the HR diagram. 
Unlike previous investigations, we make use of the synthetic 
spectrum-fitting technique for evaluation of $W_{77}$; i.e., $W_{77}$ is inversely 
computed from the oxygen abundance solution accomplishing the best fit. This approach 
is particularly effective for rapid rotators (occasionally seen in A--F giants), 
for which direct measurement of $W_{77}$ is not easy because of the contamination 
of other lines.  This is the primary purpose of this study.

Besides, as a by-product resulting from the analysis, we can study the behavior 
of microturbulence\footnote{
This is by definition the microscopic turbulent velocity dispersion, the characteristic 
scale of which is assumed to be much smaller than the photon mean-free-path. Accordingly, 
it is formally included into the Doppler width of line-opacity profile in parallel 
with the velocity of thermal motion. While the strength of a weak line (on the linear 
part of the curve of growth) is hardly affected by this parameter, that of a strong  
saturated line (on the flat part of the curve of growth) is very sensitive to it.  
} ($\xi$) over a wide parameter range of evolved A--G stars, which is considered to be 
an important parameter affecting $W_{77}$, because O~{\sc i} 7771--5 lines are strong 
and saturated in these stars. We will derive this key parameter by two (profile-based 
and abundance-based) methods; these two kinds of $\xi$ values determined by independent 
techniques would make a useful comparison.
In this context, Takeda (1992) previously studied how $\xi$ behaves 
in A--F stars of various luminosity classes, while comparing the observed $W_{77}$ data
taken from various literature with the theoretically calculated non-LTE equivalent widths.
While the global nature of $\xi$ was roughly elucidated in that paper, $\xi$ values of 
individual stars could not be discussed in terms of their dependence upon stellar parameters.
Accordingly, this is a good opportunity to challenge the task which was left undone
in Takeda (1992). Especially, since studies on the behavior of $\xi$ for A--F giants 
of luminosity class~III seem to have been barely done (see, e.g., Fig.~11 of Gray 1978), 
presumably because of the existence of rapid rotators, our analysis would make a new contribution
to this field. As such, this checking upon $\xi$ defines the second aim of this paper.

\section{Observational Data}

The targets of this study are 75 A-, F-, and G-type stars (from subgiants
through supergiants), which had already evolved off the main sequence.
The basic data of these objects are given in table 2.
(See also tableE.dat given as online material for more detailed information.) 

The observations of these objects (except for HD~20902) were carried out 
on 2012 October 6--7, 2013 March 28--29, 31, and 2013 May 21--23 by using BOES 
(Bohyunsan Observatory Echelle Spectrograph) attached to the 1.8 m reflector 
at Bohyunsan Optical Astronomy Observatory. 
Using a 2k$\times$4k CCD (pixel size of 15~$\mu$m~$\times$~15~$\mu$m), 
this echelle spectrograph enabled us to obtain spectra of wide 
wavelength coverage (from $\sim$~3600~$\rm\AA$ to 
$\sim$~9200~$\rm\AA$) at a time.
We used 200~$\mu$m fiber corresponding to the resolving power 
of $R \simeq 45000$. The integrated exposure time 
for each star was typically about ten to several tens minutes.
The reduction of the echelle spectra (bias subtraction, flat 
fielding, scattered-light correction, spectrum extraction, wavelength 
calibration, co-addition of spectra to increase S/N, and 
continuum normalization) was carried out by using the ``echelle'' package
of the software IRAF\footnote{IRAF is distributed
    by the National Optical Astronomy Observatories,
    which is operated by the Association of Universities for Research
    in Astronomy, Inc. under cooperative agreement with
    the National Science Foundation.} 
in the standard manner. For most of the targets, we 
could accomplish sufficiently high S/N ratio (typically a few hundreds) 
at the orange--red region relevant to the present study. 
Regarding HD~20902 ($\alpha$~Per), we exceptionally used the high-dispersion 
spectrum (also obtained by using BOES) published by Lee et al. (2006).

\section{Stellar Parameters}  

The parameters of 75 program stars were determined by the photometric data
(apparent magnitude $V$ and $B-V$ color) taken from the Hipparcos catalogue 
(ESA 1997) and the newly reduced Hipparcos parallaxes $\pi$ (van Leeuwen 2007).\footnote{
Although we basically adopted the new-reduction data published by van Leeuwen (2007),
the first-released Hipparcos parallaxes (ESA 1997) were exceptionally used for
8 stars as remarked in table~2 (for which the values of two catalogues are 
appreciably different because of being distant) by considering the consistency 
between the resulting $T_{\rm eff}$/$\log g$ and the spectral type/luminosity class.}
We estimated the interstellar extinction ($A_{V}$) for each star by using
Hakkila et al.'s (1997) EXTINCT program\footnote{Available at 
$\langle$http://asterisk.apod.com/library/ASCL/extinct/extinct.for$\rangle$.} 
from the galactic coordinates ($l, b$) along with the distance 
$d (\propto 1/\pi)$; and the color excess was derived as $E_{B-V} = A_{V}/3.0$.
Then, Alonso, Arribas, and Mart\'{\i}nez-Roger's (1999) Eq.(3) (for $B-V < 0.75$)
and Eq.~(4) (for $B-V \ge 0.75$) were invoked to derive $T_{\rm eff}$ from the 
reddening-corrected color $B-V$, where we assumed [Fe/H] = 0.
Further, the absolute magnitude ($M_{V}$) and bolometric luminosity ($L$) were
calculated from the extinction-corrected $V$, parallax ($\pi$), and the bolometric 
correction (B.C.) estimated with the help of Alonso et al.'s (1999) Eq. (17) and Eq. (18).

Here, we should keep in mind that Alonso et al.'s (1999) analytical formulas 
for $T_{\rm eff}$ and B.C. were derived specifically for class-III giants
(besides, $T_{\rm eff}$ formula is applicable for $T_{\rm eff} \ltsim 8000$~K,
while B.C formula is for $T_{\rm eff} \ltsim 9100~K)$.
Since we universally applied them to our program stars of A--F--G type covering 
wide range of luminosity classes (subgiants through supergiants), errors caused by 
extrapolation may be more or less expected especially for the case of supergiants.   
In order to examine this point, we compared these Alonso et al.'s formula with 
Flower's (1996) calibration for supergiants of wide $T_{\rm eff}$ range (cf. 
Table~4 therein), and found that the differences are unimportant ($\ltsim$~100--300~K 
in $T_{\rm eff}$ or $\ltsim$ a few hundredths mag in B.C.) at $B-V \gtsim 0.1$, 
while the discrepancies begin to manifestly grow once $B-V$ becomes less than 0.1 
(e.g., up to $\sim 1000$~K in $T_{\rm eff}$ and $\sim$~0.2--0.3~mag in B.C. at 
$B-V \sim 0$) in the sense that Alonso et al.'s formula underestimates both 
$T_{\rm eff}$ and $|$B.C.$|$. 
Accordingly, we may state that inadequate parameters may possibly result by our 
application of Alonso et al.'s relations for the case of A-type (especially early-A) 
supergiants. Yet, since the relevant stars of this type in our sample are distant 
and confined to the Galactic plane, their stellar parameters tend to be unreliable 
by themselves in any case, due to considerable ambiguities in parallax as well as 
interstellar extinction (see subsection~7.1).

The $\log L$ vs. $\log T_{\rm eff}$ diagram of the program stars are depicted in 
figure~1, where the evolutionary tracks (for the solar metallicity and corresponding 
stellar masses of $M \sim$~1.5--20~$M_{\odot}$) computed by Lejeune and Schaerer (2001) 
are also shown for comparison. 
Making use of the fact that these evolutionary tracks in figure~1 tend to run almost 
horizontally (i.e., $L$ is quite sensitive to $M$ but rather inert to $T_{\rm eff}$),
we assume that $M$ is a monotonic function of $L$ for any given $T_{\rm eff}$ at 
10000~K $\gtsim T_{\rm eff} \gtsim$~5000~K, where we define this function 
by numerically interpolating the light-green portions of the tracks in figure~1.   
In this way, we could evaluate $M$ (from $T_{\rm eff}$ and $L$), from which the surface 
gravity ($\log g$) was further derived by the relation $g \propto T_{\rm eff}^{4} M /L$.
The finally resulting values of $M_{V}$, $L$, $M$, $T_{\rm eff}$, and $\log g$ for 
each star are summarized in table~2; more detailed data (including the basic
photometric data, parallax, interstellar extinction, bolometric correction, etc)
are presented in the online material (tableE.dat).

These $T_{\rm eff}$ and $\log g$ values of 75 program stars determined by rather 
rough methods are compared with available literature data (taken from the SIMBAD database)
in figures~2a and 2b, respectively, where we can see a reasonable consistency for 
$T_{\rm eff}$, while some systematic trend (though not so serious) is observed for $\log g$.  
The correlations of $T_{\rm eff}$ vs. spectral type and $\log g$ vs. luminosity class
are depicted in figures~2c and 2d, respectively.  

It may be worth commenting on the accuracy of useful analytical relations.
The basic $M$ vs. $L$ relations we employed for deriving $M$ (from $L$) are displayed
in figure~3a for three values of $T_{\rm eff}$ (5000, 7500, 10000~K), which are quite
similar to each other reflecting the near-horizontal nature of evolutionary tracks
(i.e., insensitive to changes in $T_{\rm eff}$). The $L$ values for individual
stars are plotted against the resulting $M$ in figure~3b.
Since $\log L$ vs. $\log M$ relation almost follows the straight line
[$\log (L/L_{\odot}) = 0.194 + 4.07 \log (M/M_{\odot}$) according to the 
linear-regression analysis], this means that the power law of $L \propto M^{4}$ holds fairly well. 
Actually, the mutual relation between $\log g$, $M_{\rm bol}$, and $T_{\rm eff}$ for A--F stars
proposed by Takeda (1992) (later, the constant was slightly revised by Takeda and Takada-Hidai 1994) 
\begin{equation}
\log g = 0.30 M_{\rm bol} + 4 \log T_{\rm eff} - 12.05
\end{equation}
as well as the equation for the stellar mass adopted by Takeda and Takada-Hidai (1994) 
for A--F supergiants
\begin{equation}
\log (M/M_{\odot}) = [4\log (T_{\rm eff}/T_{\rm eff,\odot}) - \log (g/g_{\odot})]/3
\end{equation}
are both based on the scaling law of $L \propto M^{\alpha}$ with $\alpha = 4.0$.
In figures~3c and 3d are compared the finally adopted values of $M_{\rm bol}$ 
and $M$ with those derived by these simple analytical formulas, respectively.
We see from figure~3c that equation~(1) is sufficiently accurate (differences of $M_{\rm bol}$ 
are only $\ltsim 0.2$~mag), while application of equation (2) somewhat overestimates 
the stellar mass by $\sim$~0.05--0.1~dex ($\ltsim 20\%$).

\section{Model Atmospheres and Non-LTE Calculations}

The model atmosphere for each star was constructed by two-dimensionally 
interpolating Kurucz's (1993) ATLAS9 model grid (models with convective overshooting) 
with respect to $T_{\rm eff}$ and $\log g$ determined in section~3, 
where we exclusively applied the solar-metallicity models. 

In order to adequately incorporate the non-LTE effect, which is requisite
for the O~{\sc i} triplet lines under study, we carried out non-LTE calculations 
for oxygen on an extensive grid of solar-metallicity model atmospheres resulting 
from combinations of eleven 
$T_{\rm eff}$ values (5500, 6000, 6500, 7000, 7500, 8000, 8000, 8500, 
9000, 9500, 10000~K) and six $\log g$ values (1.5, 2.0, 2.5, 3.0, 3.5, 4.0),\footnote{
Since ATLAS9 model grid does not include $\log g= 1.5$ models at  
$T_{\rm eff} \ge 9500$~K presumably because of an instability problem,   
models of only five gravities ($\log g$ = 2.0, 2.5, 3.0, 3.5, and 4.0) 
could be used for $T_{\rm eff}$ = 9500 and 10000~K.} and eleven $\xi$ values (0, 1, 2, 3, 4, 
5, 6, 7, 8, 9, 10~km~s$^{-1}$), which almost cover the parameter ranges of our program stars.
See Takeda (2003) and the references therein for details
of the calculation procedures. Regarding the treatment of collisional rates
with neutral hydrogen atoms, we adopted the conventional treatment without any corrections ($k=1$). 
The non-LTE departure coefficients [$b(\tau)$] applied to each star were then derived 
by interpolating this grid in terms of $T_{\rm eff}$ and $\log g$, 
as was done for model atmospheres. Since departure coefficients are not sensitive
to a change of $\xi$, an appropriate dataset of the grid near to the relevant $\xi$
value was employed.

\section{Spectrum Fitting of O I 7771--5} 

Since the spectra of A--F--G stars in the 7765--7785~\AA\ region (comprising
 O~{\sc i} 7771--5 triplet lines and Fe~{\sc i} 7781 line) depend on
various factors (O and Fe abundances, microturbulence, macroturbulence/rotational 
velocity, radial velocity, etc.), we may obtain information of these
parameters by a careful analysis based on the spectrum synthesis technique, 
as previously done by Takeda and Sadakane (1997) for A--F stars in the Hyades cluster. 

Accordingly, we searched for the solutions for the oxygen abundance ($A$(O)), 
Fe abundance (A(Fe)), microturbulence ($\xi_{\rm p}$),\footnote{Hereinafter, we refer
to this parameter as $\xi_{\rm p}$ (``p'' means ``profile'') because this is the 
microturbulence derived from line profiles.} macrobroadening velocity 
($v_{\rm M}$), and radial velocity ($V_{\rm rad}$) 
which accomplish the best fit (minimizing $O-C$ residuals) 
between the theoretical and observed spectrum in the 7765--7785~\AA\ region, 
by applying the automatic fitting algorithm described in Takeda (1995).
Regarding the macrobroadening function (to be convolved with the intrinsic
profile), we adopted the classical rotational broadening function (see, 
e.g., Gray 2005) with the limb-darkening coefficient of $\epsilon = 0.5$.
Therefore, $v_{\rm M}$ may be regarded as equivalent to $v_{\rm e}\sin i$ (projected
rotational velocity) if macroturbulence is negligible. Generally, if $v_{\rm M}$ is larger 
than several tens km~s$^{-1}$, we may safely consider $v_{\rm M} \simeq v_{\rm e}\sin i$;
otherwise, appreciable contribution of macroturbulence to $v_{\rm M}$ (in addition to
$v_{\rm e}\sin i$) is quite probable. As to the atomic data  of spectral lines, 
we exclusively consulted the compilation by Kurucz and Bell (1995). 
The data for the three O~{\sc i} 7771--5 lines are summarized in table~3. Although 
all atomic lines given in their compilation for this wavelength region were 
included in the spectrum synthesis, the abundances of elements other than 
O and Fe were fixed at the solar abundances.  

The convergence of the solutions was satisfactory for most cases.\footnote{
Practically, some tricks are needed in order to accomplish successful convergence 
for all parameters. For example, iteration should be 
done by fixing $\xi$ at the first round, followed by the second round of
iteration (starting from converged solution in the first round) where
$\xi$ is allowed to vary toward convergence.} Nevertheless, we sometimes encountered 
instability or divergence of solutions; in such cases, we had to fix the 
relevant parameter at an appropriate value. The resulting velocity parameters 
($v_{\rm M}$ and $\xi_{\rm p}$) for each star are summarized in table~2. 
It is demonstrated in figure~4 that the theoretical spectra for the converged 
solutions properly fit with the observed spectra.

We then inversely computed the equivalent width ($W_{77}$) for the whole 
O~{\sc i} 7771--5 triplet by using the finally converged solutions of $A$(O) 
and $\xi_{\rm p}$ along with the same atomic data and model atmosphere 
used in the fitting analysis. For this purpose, we used Kurucz's (1993) WIDTH9
program, which was considerably modified in many respects (e.g., to incorporate 
the non-LTE effect, to allow complex feature comprising a multiple of line 
components, etc.). The $W_{77}$ values finally obtained as such, based on which 
we will discuss the luminosity effect of O~{\sc i} 7771--5 lines, are given in table~2.

\section{Abundance-Based Microturbulence} 

Although we derived the microturbulence ($\xi_{\rm p}$) from the profile of 
O~{\sc i} 7771--5 triplet in section~5, this is not the usual approach.
The conventional method of determining this parameter is to require the
consistency of abundances derived from strong and weak lines, which makes 
use of the fact that the former abundance is much more $\xi$-sensitive 
than the latter. It is meaningful to determine also this $\xi_{\rm a}$\footnote{
Hereinafter, we refer to this parameter as $\xi_{\rm a}$ (``a'' means ``abundance'') 
because this is the microturbulence derived by requiring the abundance consistency
between lines of different strengths.} and see how these two kinds of 
microturbulence are compared with each other. For this purpose, we use
O~{\sc i} 6155--8 lines (3p~$^{5}$P--4d~$^{5}$D$^{\rm o}$, multiplet 10),
which are much weaker than O~{\sc i} 7771--5 lines and thus suitable.
Although the orange region where these lines situate is more crowded 
with other spectral lines compared to the case of O~{\sc i} 7771--5,
oxygen abundance can be determined (even for rapid rotators) by applying 
the spectrum-synthesis technique, as recently done by Takeda, Hashimoto, 
and Honda (2017) for their study of F--G type stars in the Pleiades cluster.

In almost the same manner as done in section~5, we conducted a spectrum-fitting
analysis in the 6143--6168~$\rm\AA$ region, while changing the abundances of O, Na,
Si, Ca, and Fe to search for the best-fit solution (microturbulence was fixed at 
$\xi_{\rm p}$). The finally accomplished fit between the theoretical and observed
spectra are demonstrated in figure~5. Then, the equivalent width ($W_{61}$) 
corresponding to whole O~{\sc i} 6155--8 was evaluated from the converged 
solutions of $A$(O) and $\xi_{\rm p}$ by integrating the synthesized 
spectrum of 9 component lines (cf. table~3) as done in section~5 for $W_{77}$.

We computed the non-LTE oxygen abundances ($A^{\rm N}_{77}$ and $A^{\rm N}_{61}$)
(and also the LTE abundances $A^{\rm L}$ as well as the non-LTE correction 
$\Delta \equiv A^{\rm N} - A^{\rm L}$)
from such established $W_{77}$ and  $W_{61}$ for each star while progressively 
changing the $\xi$ values. Based on this set of abundances, $\xi_{a}$ may be 
defined as the $\xi$ value satisfying  $A^{\rm N}_{77} = A^{\rm N}_{61}$. 
Although this attempt was not always successful, we could determine $\xi_{\rm a}$ 
for 62 stars (cf. table~2), which will be discussed in subsection 7.4 in comparison
with $\xi_{\rm p}$. Figure~6 shows the $T_{\rm eff}$-dependence of equivalent widths 
($W$), microturbulence ($\xi_{\rm a}$), non-LTE abundances ($A^{\rm N}$) as well as 
non-LTE corrections ($\Delta$) corresponding to $\xi_{\rm a}$.
The complete data of $W$, $A$(O), and $\Delta$ for both line features are presented
in tableE.dat of the online material.  

\section{Discussion}

\subsection{Accuracy of absolute magnitude and equivalent width} 

Since the main purpose of this paper concerns the empirical relation between $M_{V}$ 
and $W_{77}$, it may be appropriate here to mention the accuracies of these quantities 
(cf. table~2) which we derived in section~3 and section~5.

Regarding $M_{V}$, two factors are involved in its uncertainty: (i) error ($\sigma_{\pi}$)
in the Hipparcos parallax ($\pi$), which contributes to an error in $M_{V}$ by the relation
$\delta M_{V}^{\pi} \simeq 5 \log (1 + \sigma_{\pi}/\pi)$, and (ii) error in the adopted 
interstellar extinction ($\delta M_{V}^{A} = \sigma_{A}$), which is estimated by 
the standard deviation of $A_{V}$ resulting from the EXTINCT program. 
Then, the total error of $M_{V}$ may be evaluated by the square-sum-root of these two as
$\delta M_{V}^{\pi+A} \equiv \sqrt{(\delta M_{V}^{\pi})^{2}+(\delta M_{V}^{A})^{2}}$.
We plot such evaluated $\delta M_{V}^{\pi}$, $\delta M_{V}^{A}$, and $\delta M_{V}^{\pi+A}$
for eah star against $M_{V}$ in figure~7a, 7b, and 7c, respectively.
We can see the following characteristics from these figures.\\
--- Both $\delta M_{V}^{\pi}$ and $\delta M_{V}^{A}$ tend to increase as $M_{V}$ becomes brighter,
which simply reflects the fact that brighter $M_{V}$ stars are generally more distant
and thus suffer larger errors in $\pi$ as well as in $A_{V}$.\\
--- There is a tendency that the extent of $\delta M_{V}^{\pi}$ is larger 
than that of $\delta M_{V}^{A}$ (especially for brighter $M_{V}$ stars), which means that 
parallax error is comparatively more important in affecting the total uncertainty of $M_{V}$.\\
--- According to figure~7c, the error of $M_{V}$ is typically several tenths of magnitude 
for giants ($M_{V} \gtsim -2$), while it tends to suddenly grow with luminosity  
at $-2 \ltsim M_{V}$ and even attains as much as $\sim 1$~mag at $M_{V} \sim -5$.\\
--- These intrinsically luminous supergiants (around $M_{V} \sim -5$) with considerable
$M_{V}$ errors have luminosities around $\log (L/L_{\odot}) \sim $4--5. Since such
stars tend to be of rather high $T_{\rm eff}$ ($\gtsim 8000$~K) as seen from figure~1, 
we should bear in mind that our $M_{V}$ values of A-type supergiants may suffer rather 
large uncertaities.  

On the other hand, we may expect that $W_{77}$ could be established with a rather 
high precision, because it was derived by applying the spectrum-fitting technique, 
where a sufficiently good fit could be accomplished by the theoretical synthesized 
spectrum regardless of how complex the line profile is.

Regarding the error of equivalent width ($W$) stemming from the spectrum noise, 
we may invoked the formula derived by Cayrel (1988)
\begin{equation}
\delta W \simeq 1.6 (w \delta x)^{1/2} \epsilon,
\end{equation}
where $\delta x$ is the pixel size (or sampling step), $w$ is the
full-width at half maximum, and $\epsilon \equiv ({\rm S/N})^{-1}$.
Putting $w \simeq 7773  v_{\rm M}/c$ ($w$ is in \AA\ and $c$ is the velocity of light), 
$\delta x \simeq 0.04$\AA, and S/N~$\sim 200$ (typical value),
we obtain $\delta W \sim (1/1000)\sqrt{v_{\rm M}/10}$ (where $\delta W$ is in \AA\ 
and $v_{\rm M}$ is in km~s$^{-1}$).
In case of small $v_{\rm M}$ where the three component lines are clearly split,
we may regard $W \simeq W_{77}/3$, while $W = W_{77}$ for the case of merged triplet 
with large $v_{\rm M}$. Considering four representative cases resulting from combinations
of $v_{\rm M}$ = 10 and 100~km~s$^{-1}$ and $W_{77}$ = 0.5 and 1~\AA, we can conclude
that $\delta W/W$ is $<$1\% in any event, which is negligibly insignificant.

For reference, we also compare our $M_{V}$ and $W_{77}$ values with those published by 
Arellano Ferro et al. (2003; 6 stars in common; mean equivalent widths designated as
``$W_{74,0}$'' in their Table~2 were adopted for four Cepheids) as well as Kovtyukh et al. 
(2012; 11 stars in common; ``literature $M_{V}$'' given in their Table~1 were used) 
in figure~7d and figure~7e, respectively, where we can see a reasonable consistency 
without any systematic difference. 

\subsection{Observed behavior of $W$(O I 7771--5)}  

We now discuss how $W_{77}$ depends on the stellar parameters, especially 
in terms of the absolute magnitude. In figure~8 are plotted the resulting $W_{77}$
values of 75 program stars against $T_{\rm eff}$, $\log g$, $v_{\rm M}$, and $M_{V}$.

In figure~8a, the triplet line strengths for 46 A-type dwarfs (which were inversely computed 
from the O abundance results of Takeda et al. 2008a) as well as those of 160 FGK dwarfs
(which were obtained by summing-up the observed equivalent widths of three lines 
published by Takeda \& Honda 2005) are also shown for comparison (note that oxygen-deficient 
stars are included in these samples of main-sequence stars).
Immediately noticeable from figure~7a is the increasing tendency toward higher $T_{\rm eff}$ 
(especially from G to F), which is 8ctually an expected trend for such high-excitation lines
(cf. figure~9a).  The fact that $W_{77}$ values of almost all evolved stars of our sample 
surpass those of main-sequence stars at any given $T_{\rm eff}$ evidently indicates
that O~{\sc i} 7771--5 triplet generally strengthens as a result of stellar evolution. 

Regarding the $W_{77}$ vs. $v_{\rm M}$ plot shown in figure~8c, we see a correlation 
that $W_{77}$ steeply grows with an increase in $v_{\rm M}$ for $\log g < 2.5$ stars 
(open circles) at comparatively low $v_{\rm M}$ values being less than several tens km~s$^{-1}$.
Recalling that appreciable contribution of macroturbulence (or almost dominated
by the macroturbulence) is expected for $v_{\rm M}$ of such stars (cf. section~5),
this prominent growth of $W_{77}$ with $v_{\rm M}$ is attributed to the increase of $\xi$
(compare figure~6a with figure~6b), because micro- and macro-turbulence are likely 
to be closely connected. Meanwhile, no clear $v_{\rm M}$-dependence is observed in
$W_{77}$ for stars with larger values of $v_{\rm M}$, where $v_{\rm M}$ is essentially
equivalent to $v_{\rm e}\sin i$.

Since surface gravity and absolute magnitude are intimately related with each other
as indicated by equation~(1), it is natural that $W_{77}$ vs. $\log g$ plot (figure~8b)
and $W_{77}$ vs. $M_{V}$ plot (figure~8d) look quite similar.
It is figure~8d that reveals the nature of the luminosity effect for O~{\sc i} 7771--5 
in evolved A--G stars. We note the following characteristics from this figure.
\begin{itemize}
\item
We can observe that lower gravity stars of $\log g < 2.5$ (open circles)
and higher gravity stars of $\log g > 2.5$ (filled circles) show apparently different 
behaviors of $W_{77}$. 
\item
Remarkably, although the various empirical $W_{77}$ vs. $M_{V}$ formulas proposed 
so far (gray lines) are more or less consistent with the distribution of former lower-gravity 
group (supergiants) at the wide $M_{V}$ range ($1 \gtsim M_{V} \gtsim -8$), they never 
represent that of the latter higher-gravity group (subgiants, giants).
\item
We should be careful for the stars of $0 \gtsim M_{V} \gtsim -5$, in which 
these two groups are mixed; i.e., lower $T_{\rm eff}$/lower $\log g (<2.5)$ stars
and higher $T_{\rm eff}$/higher $\log g (> 2.5)$ stars. Again, the published formulas
summarized in table~1 are relevant only for the former (F--G supergiants) while
not for the latter (A-type giants).
\item
The trend of $W_{77}$ vs. $M_{V}$ for the higher-gravity ($\log g >2.5$) group 
(red lines depicted in figure~8d) is roughly represented by a steep increase
of $W_{77}$ from $M_{V} \sim +3$ to $\sim 0$, followed by only gradual increase
from $M_{V} \sim 0$ to $\sim -4$. 
\end{itemize}

Accordingly, regarding the question about the applicability limit of published $W_{77}$ vs. $M_{V}$ 
formulas, which was raised in section~1 and motivated this study, our answer is as follows:\\
--- (1) These analytical relations may be safely applied to intrinsically 
bright supergiants ($-5 \gtsim M_{V} \gtsim -10$), which have low gravities of 
$\log g \ltsim 2.5$ in any case. However, high-precison would not be expected 
in empirically estimating $M_{V}$ from $W_{77}$, because appreciable dispersion 
amounting to $\ltsim 2$~mag in $M_{V}$ is observed in those proposed formulas 
(cf. gray lines in figure~8d).
Moreover, our results (larger open circles in figure~8d) show some systematic 
difference (our $M_{V}$ values show a rather large scatter but tend to be 
dimmer by $\sim$~1--2~mag on the average) as compared to these published relations. 
This manifestly reflects the general difficulty of $M_{V}$ determination for 
such bright supergiants suffering large unertainties in $A_{V}$ as well as 
in $\pi$ because they are distantly located in the Galactic plane 
(note that the error in $M_{V}$ begins to prominently grow from several tenths mag
at $M_{V} \sim -2$ to a considerable level of $\sim$~1~mag or even larger 
at $M_{V} \sim -5$; see figure~7c).\\
--- (2) As to giants or bright giants of $0 \gtsim M_{V} \gtsim -5$, the situation 
is complicated because $\log g \ltsim 2.5$ group and $\log g \gtsim 2.5$ group 
show markedly different trends. While the simple conventional relations
(such as that proposed by Arellano Ferro et al. 2003, which well fits our
results of small open circles in figure~8d) may still be usable for the former 
(typically F--G supergiants), they are no more valid for the latter (typically A giants).
If $W_{77}$ values of these mixed samples altogether are to be explained by analytical 
relations, one would have to invoke complex formulas depending not only on $M_{V}$ 
but also on other stellar parameters.\\
--- (3) In short, regarding the luminosity effect of O~{\sc i} 7771--5 triplet,   
application (or devising) of useful analytical $W_{77}$ vs. $M_{V}$ formulas had better 
be confined to only bright supergiants of $-5 \gtsim M_{V} \gtsim -10$. We should bear 
in mind, however, that such relations are quantitatively of limited accuracy, because 
of the enormous difficulty in precisely calibrating $M_{V}$ of such generally distant 
stars suffering large interstellar extinction. 

\subsection{Comparison with theoretical calculations}  

Now that the observed trend of O~{\sc i} 7771--5 line strength in terms of stellar 
parameters has been elucidated, it is meaningful to check how this is compared with 
theoretically computed values of $W_{77}$. In our line-formation calculation, the non-LTE effect was
taken into account based on plane-parallel model atmospheres, in which microturbulence was
assumed to be depth-independent. We consider this treatment/assumption is practically sufficient 
(even if not complete), since the behaviors of non-LTE effect and (especially) microturbulence,
both of which tend to increase toward lower-density atmospheres of higher-luminosity stars,
should be the main factors that dominate this luminosity effect. 

Although Przybilla et al. (2000) suggested that non-classical nature of stellar atmospheres 
(spherical extension effect, outflow velocity field) may be involved with the considerable 
intensification of O~{\sc i} 7771--5 lines in A-type supergiants, their argument does not seem 
to be based on a firm evidence. Regarding normal high-mass supergiants with static atmospheres, 
the sphericity effect is not significant and application of plane-parallel model atmospheres 
is not bad in the practical sense, since the thickness of atmosphere ($d$) is still 
considerably smaller than the radius ($R$).\footnote{
Since $R$ and $d$ can be expressed as $R \propto T_{\rm eff}^{-2} L^{1/2}$ 
and $d \propto g^{-1} \propto R^{2} M^{-1} \propto T_{\rm eff}^{-4}L M^{-1}$, 
we have $d/R \propto T_{\rm eff}^{-2} L^{1/2} M^{-1}$.
Accordingly, as $d/R$ is inversely proportional to $M$, the sphericity effect is not 
important for ordinary supergiants of large $M$, since extended $d$ due to low gravity 
tends to be compensated by expanded $R$. We should note, however, that this effect can be 
significant for low-mass and low-gravity stars (classified superficially as supergiants), 
such as the case of post-AGB stars.} 
Admittedly, the situation must be different for the case of expanding atmospheres with 
significant mass loss. However, it seems unlikely that such an unusual phenomenon is commonly 
relevant for A--F--G giants and supergiants under study.

We computed $W_{77}$ (non-LTE and LTE, solar oxygen abundance, corresponding to 
3 $\xi$ values of 2, 5, and 10~km~s$^{-1}$) on a grid of models resulting from combinations 
of 11 values of $T_{\rm eff}$ (5000, 5500, $\cdots$ 10000~K) and 6 values of $\log g$  
(1.5, 2.0, $\cdots$ 4.0), as done for computing the grid of non-LTE departure coefficients (section~4). 
Figures~9a and 9b show the run of $W_{77}^{\rm N}$ as well as $W_{77}^{\rm L}$ with $T_{\rm eff}$ 
(for $\log g = 2.5$ case) and that with $\log g$ (for $T_{\rm eff}$ = 7500~K case), respectively.
It can be observed that $W_{77}^{\rm N}$ increases as $T_{\rm eff}$ becomes higher up to $\sim 8000$~K
where it attains a broad peak. We also see that the line is conspicuously intensified by the non-LTE 
effect and by increasing $\xi$, and the non-LTE strengthening grows as $\log g$ is decreased.
(especially for A-type stars).

The theoretical $W_{77}^{\rm N}$ vs. $M_{V}$ plots for three different $T_{\rm eff}$ groups 
are depicted in figures~9c, 9d, ad 9e, where the empirical trends are also shown by solid lines. 
We can recognize from these figures that these relations computed with 
given (constant) $\xi$ values generally fail to explain the observed data; i.e., 
the growth of observed $W_{77}$ with a decrease in $M_{V}$ (increasing luminosity)
is much steeper than the theoretical expectation. This means that an increase of $\xi$
as a star becomes brighter is essential to explain the observed $M_{V}$-dependence
of $W_{77}$. We should note, however, that $\log g \gtsim 2.5$ stars at $0 \gtsim M_{V} \gtsim -4$
may belong to an exceptional case, since the observed gradual trend (red line) almost match
the theoretical gradient. This indicates that $\xi$ values of these stars may not show
any significant dependence on $M_{V}$ (see also subsection~7.4).

\subsection{Trend of microturbulence}  

Finally, we discuss how the microturbulence depends on stellar parameters
based on the two kinds of $\xi$ results derived in section~5 ($\xi_{\rm p}$)
and section~6 ($\xi_{\rm a}$). These ($\xi_{\rm p}$, $\xi_{\rm a}$) are plotted against
$T_{\rm eff}$, $\log g$, and $v_{\rm M}$ in figures 10a,a$'$, 10b,b$'$, and 10c,c$'$,
respectively. The correlation of $\xi_{\rm p}$ and $\xi_{\rm a}$ is shown in figure~10d.

It can be seen from figures~10d that $\xi_{\rm p}$ and $\xi_{\rm a}$ are
more or less consistent with each other, despite that these two were derived
by different approaches. Actually, the distribution of $\xi_{\rm p}$ in figure~10a,b,c
(left panels) and that of figure~10a$'$,b$'$,c$'$ (right panels) appear almost similar.
A notable exception is that $\xi_{\rm p}$ tends to be appreciably larger than 
$\xi_{\rm a}$ for four A-type supergiants (large open circles in figure~10d)
where O~{\sc i} 7771--5 feature is very strong  ($W_{77} > 1$~\AA).
We suspect that this difference is related to the depth-dependence of $\xi$,
because this parameter is likely to increase with atmospheric height in supergiants
(see, Appendix B of Takeda \& Takada-Hidai 1994). That is, since $\xi_{\rm p}$ is 
mainly determined by core profile of strong feature where the formation depth 
is higher, it tends to reflect the condition of upper atmosphere (where $\xi$ is 
larger than the deeper layer), which eventually leads to $\xi_{\rm p} > \xi_{\rm a}$.
Actually, a similar phenomenon was previously observed by Takeda et al. (1996; cf. 
Sect.~5.1.1 therein) for the Sun and Procyon (though the inequality relation was 
reversed in that case of dwarfs, indicating a decrease of $\xi$ with height).  
 
We can see from figure~10a,a$'$ that the $\xi$ vs. $T_{\rm eff}$ relations 
for A dwarfs and FGK dwarfs (solid lines) almost coincide with the lower envelope 
of the $\xi$ distribution derived for our program stars. 
Accordingly, a growing of $\xi$ is a natural tendency of evolved A--G stars, 
which generally occurs as a star leaves off the main sequence. 

Figure~10b,b$'$ shows how $\xi$ depends on $\log g$, where we can observe that $\xi$ 
roughly grows with a decrease of $\log g$. This suggests that turbulence tends to be 
enhanced as the atmospheric density is lowered, which is intuitively reasonable.\footnote{
This tendency is seen also in late G through early K giants (evolved stars with 
lower $T_{\rm eff}$), as shown in  Fig.~1c of Takeda, Sato, and Murata (2008b).}
However, this trend is not simply monotonic but rather intricate.
For example, $\xi$ values of A supergiants (large open circles) are distinctly 
larger than those of F--G supergiants (small open circles) at the same $\log g \sim 2$.
We also see that, for stars of $3.0 \gtsim \log g \gtsim 2.5$ (filled circles), 
$\xi$ ($\sim$~3--4~km~s$^{-1}$) is rather inert to a change of $\log g$ . 
Since stars of this $\log g$ range almost correspond to a $M_{V}$ range 
between $\sim 0$ and $\sim -5$ according to equation (1), this indicates the existence 
of stagnant trend in $\xi$ already suggested in subsection 7.3.  
Further noteworthy is the sign of $\xi$ hump (up to $\ltsim 10$~km~s$^{-1}$) for stars of 
$\log g \sim 3$ (these have $T_{\rm eff} \sim 6000$~K), which is not easy to interpret.    

Regarding the $\xi$ vs. $v_{\rm M}$ relation shown in figure~10c,c$'$,
what we can state is almost similar to what we already discussed about 
the $v_{\rm M}$-dependence of $W_{77}$ in subsection~7.2.
The steep increase of $\xi$ with $v_{\rm M}$ at 0~km~s$^{-1} \ltsim v_{\rm M} \ltsim 40$~km~s$^{-1}$
for $\log g < 2.5$ stars must be due to the close correlation between macroturbulence and
microturbulence, since macroturbulence significantly contributes to (or dominates) 
$v_{\rm M}$ in this range of comparatively small $v_{\rm M}$. For large $v_{\rm M}$ region 
(more than several tens km~s$^{-1}$) where $v_{\rm M} \simeq v_{\rm e}\sin i$ almost holds, 
we can not observe any meaningful dependence of $\xi$ upon $v_{\rm M}$. Accordingly, 
microturbulence is not likely to be explicitly affected by stellar rotation.

\section{Conclusion}

Although the strength of O~{\sc i} 7771--5 feature is known to show a 
luminosity effect and may be usable for empirical evaluation of absolute 
magnitude, our understanding on its behavior is still insufficient.
Especially, the validity and applicability limit of various analytical relations
proposed so far (which are not necessarily consistent with each other)
has yet to be clarified, for which comprehensive study on the parameter 
dependence of $W_{77}$ for evolved stars in general would be needed.

With an aim to shed light on these points, we carried out an extensive non-LTE 
spectrum-fitting analysis of O~{\sc i} 7771--5 lines for unbiased sample of 
75 evolved A-, F,- and G-type stars of various luminosity classes (subgiants, 
giants, and supergiants) including rapid rotators, from which $W_{77}$ 
was derived for each star. Besides, as a by-product of analysis, we determined
the microturbulence (which plays a significant role in controlling $W_{77}$) 
by two different approaches (profile-based method and abundance-based method) 
for each star, because its behavior of evolved stars across the HR diagram 
is not yet well understood.

We confirmed that the resulting $W_{77}$ values of the program stars tend to
increase as $M_{V}$ becomes more luminous. However, the behavior of $W_{77}$ for 
the whole sample is too complicated to be described by a simple relation. 
Specifically, distinctly different trends of $W_{77}$ were found for 
the lower-gravity ($\log g \ltsim 2.5$) group and the higher-gravity 
($\log g \gtsim 2.5$) group; and the simple $M_{V}$ vs. $W_{77}$ formulas 
proposed by past studies are applicable only to the former group,
but not to the latter group which shows a totally different tendency.
Since these two groups overlap at $0 \gtsim M_{V} \gtsim -5$ (i.e., F--G 
supergiants of the former group and A-type giants of the latter group),
special care should be taken in using $W_{77}$ of stars in this $M_{V}$ range., 
It is thus recommended to confine only to supergiants of $-5 \gtsim M_{V} \gtsim -10$, 
if one wants to safely make use of the luminosity effect of O~{\sc i} 7771--5 lines. 

Concerning our question raised in section~1 (see the second paragraph therein), 
which motivated this investigation, our conclusion is that the relation between $M_{V}$ and 
$W_{77}$ over a wide $M_{V}$ range (covering giants and supergiants from $\sim 0$~mag 
to $\sim -9$~mag) can not be represented by such a simple analytical formula as derived by 
Arellano Ferro et al. (2003). Presumably, their sample stars (which they used for 
calibration to derive the formula) were not so sufficiently diversified as to detect 
the dispersion of $W_{77}$ for giants of $M_{V} \gtsim -5$.  

Regarding the behavior of microturbulence, it roughly shows an increasing tendency 
with a decrease in surface gravity, which we believe to be the primary cause of 
the luminosity-dependence of $W_{77}$. However, the trend is not monotonic but rather 
intricate depending on the stellar type and evolutionary stage (e.g., hump around 
$\log g \sim 3$, stagnation at $\log g \sim$~3--3.5, or discontinuously large increase 
up to $\gtsim 10$~km~s$^{-1}$ for luminous A-type supergiants). These results may 
be used as observational constraints for investigating the nature of velocity fields
in the atmosphere of evolved A--F--G stars.

\bigskip
The first author (Y. T.) heartily thanks Dr. J. Hakkila for informing how to get his EXTINCT code. 
This research has made use of the SIMBAD database, operated at CDS, Strasbourg, France.
Data reduction was in part carried out by using the common-use data analysis 
computer system at the Astronomy Data Center (ADC) of the National Astronomical 
Observatory of Japan.


\newpage
\onecolumn

\setcounter{figure}{0}
\begin{figure}
\begin{center}
  \FigureFile(100mm,100mm){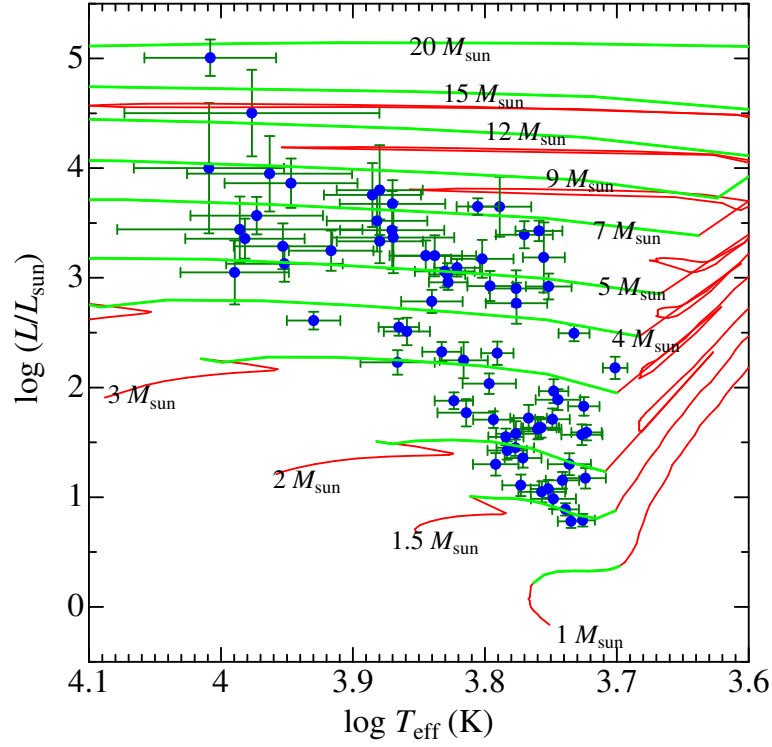}
\end{center}
\caption{
Our 75 program stars plotted on the theoretical HR diagram ($\log (L/L_{\odot})$ 
vs. $\log T_{\rm eff}$), where $T_{\rm eff}$ was determined from the $B-V$ color 
(corrected for interstellar reddening) while $L$ was evaluated from visual magnitude 
(corrected for interstellar extinction), Hipparcos parallax, and bolometric correction. 
The error bars in $T_{\rm eff}$ are due to ambiguities of interstellar reddening, 
while those in $L$ are estimated by combining the uncertainties of interstellar 
extinction and of Hipparcos parallax.
Theoretical solar-metallicity tracks computed by Lejeune and Schaerer (2001) are 
also depicted for 11 different masses  (1, 1.5, 2, 3, 4, 5, 7, 9, 12, 15, and 
20~$M_{\odot}$) for comparison. The nearly horizontal parts of the tracks 
(corresponding to the shell-hydrogen burning phase) colored in light-green 
were used to estimate $M$ for given $L$ and $T_{\rm eff}$ by interpolation.
}
\label{fig:1}
\end{figure}

\setcounter{figure}{1}
\begin{figure}
\begin{center}
  \FigureFile(120mm,150mm){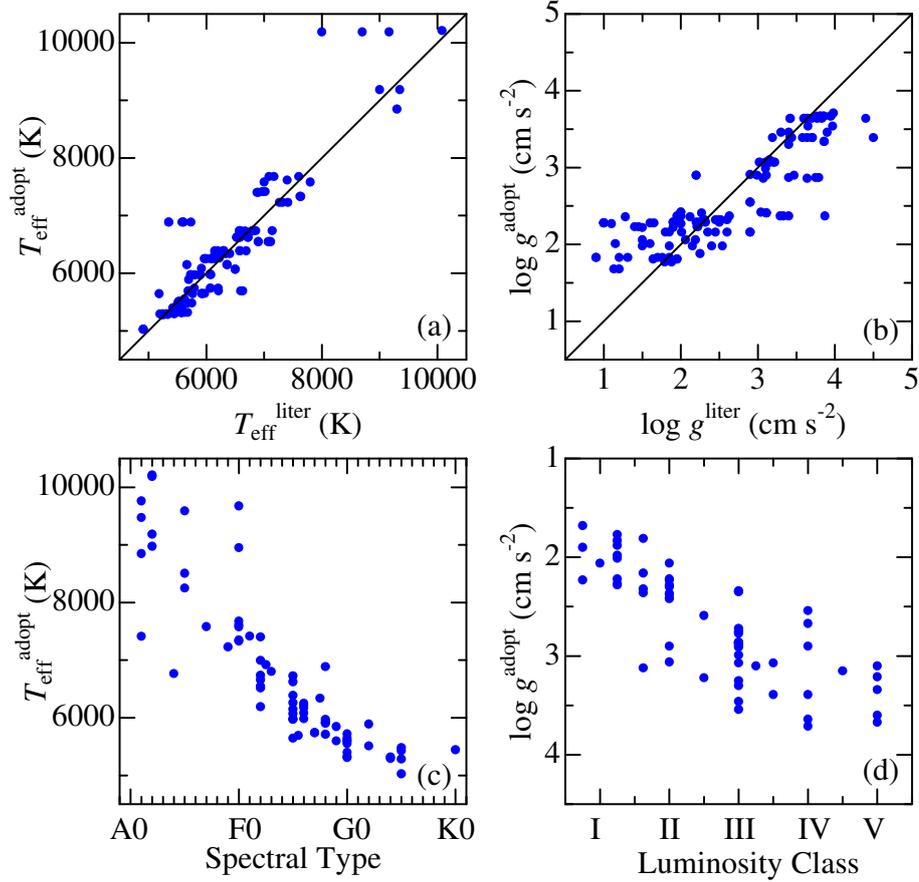}
\end{center}
\caption{
Panels (a) and (b) show the comparison of $T_{\rm eff}$ and $\log g$ adopted in this study 
with those of literature data taken from the simbad database (available for 
$\sim 40$ stars, though the number of the plotted points is larger because 
several published data tend to be attached for each star).
In panels (c) and (d) are depicted the correlations of $T_{\rm eff}$ vs. spectral type 
and $\log g$ vs. luminosity class, respectively, where subtypes were tentatively 
digitized (such as like Ia $\rightarrow 0.75$ and Ib $\rightarrow 1.25$). 

}
\label{fig:2}
\end{figure}

\setcounter{figure}{2}
\begin{figure}
\begin{center}
  \FigureFile(120mm,150mm){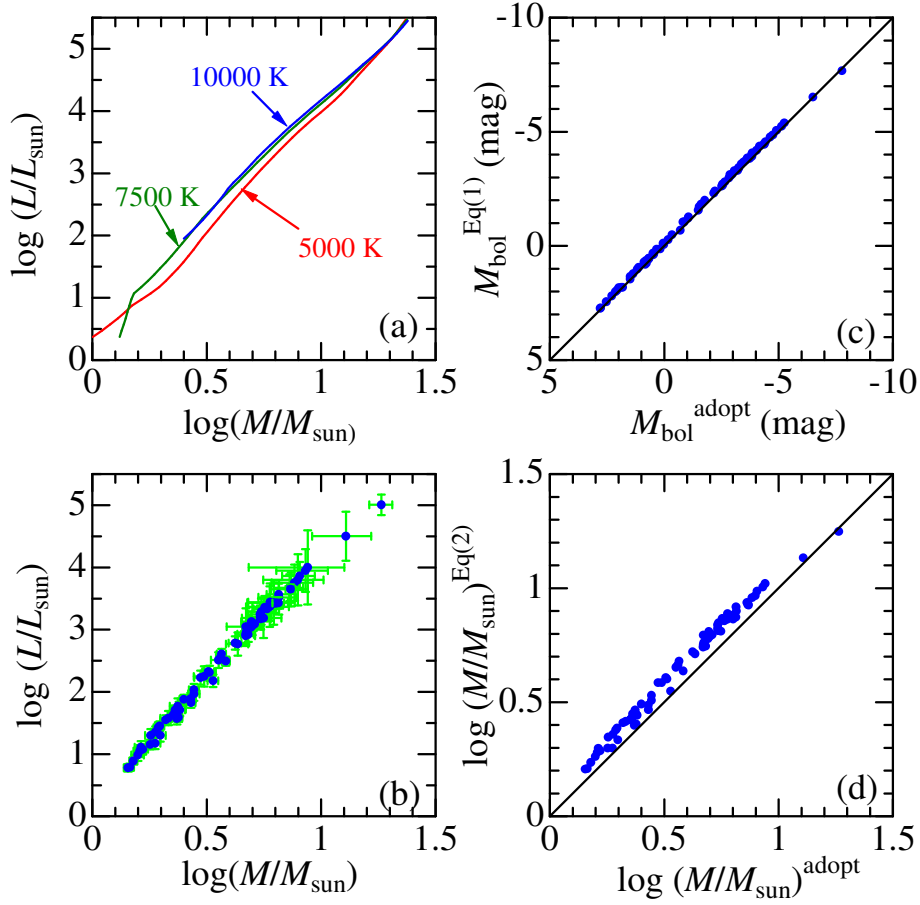}
\end{center}
\caption{(a) $L$ vs. $M$ relation (for given $T_{\rm eff}$) derived from the near-horizontal 
parts (colored in light-green in figure~1) of the evolutionary tracks. Depicted here 
are the cases of $T_{\rm eff} = 5000$, 7500, and 10000~K, where we can see that almost 
the same relation holds irrespective of $T_{\rm eff}$.
(b) Mass ($M$) values of the program stars (which were derived by using these 
$M = f(L, T_{\rm eff})$ relations by interpolation) plotted against luminosity ($L$).
The attached error bars are due to the ambiguities in $L$ (see the caption in figure~1).
(c) Correlation of $M_{\rm bol}$ (absolute bolometric magnitude) values 
derived in this study with those computed by equation~(1). (d) Correlation of $M$ (mass) 
values derived in this study with those computed by equation~(2).
}
\label{fig:3}
\end{figure}

\setcounter{figure}{3}
\begin{figure}
\begin{center}
  \FigureFile(140mm,200mm){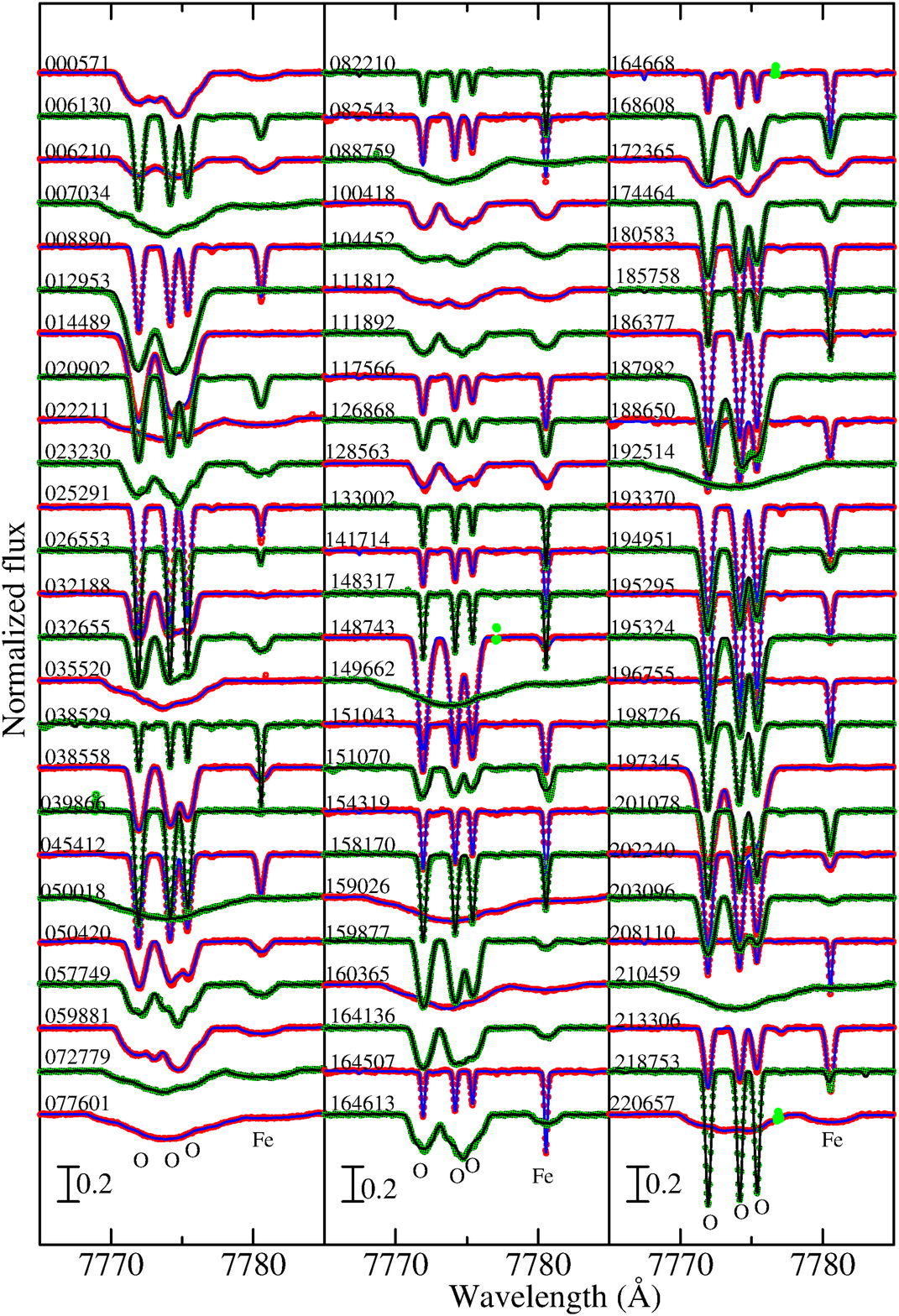}
\end{center}
\caption{
Synthetic spectrum fitting at the 7765--7785~$\rm\AA$ region.
The best-fit theoretical spectra are shown by solid lines and the observed data 
are plotted by symbols (while those masked/disregarded in the fitting are 
highlighted in light-green).  
In each panel, the spectra are arranged in the order of star's HD number 
(indicated in the figure) as in table 2, and an offset of 0.2 is applied to each 
spectrum relative to the adjacent one. 
The wavelength scale is adjusted to the laboratory frame
by correcting the radial-velocity shift.
}
\label{fig:4}
\end{figure}

\setcounter{figure}{4}
\begin{figure}
\begin{center}
  \FigureFile(140mm,200mm){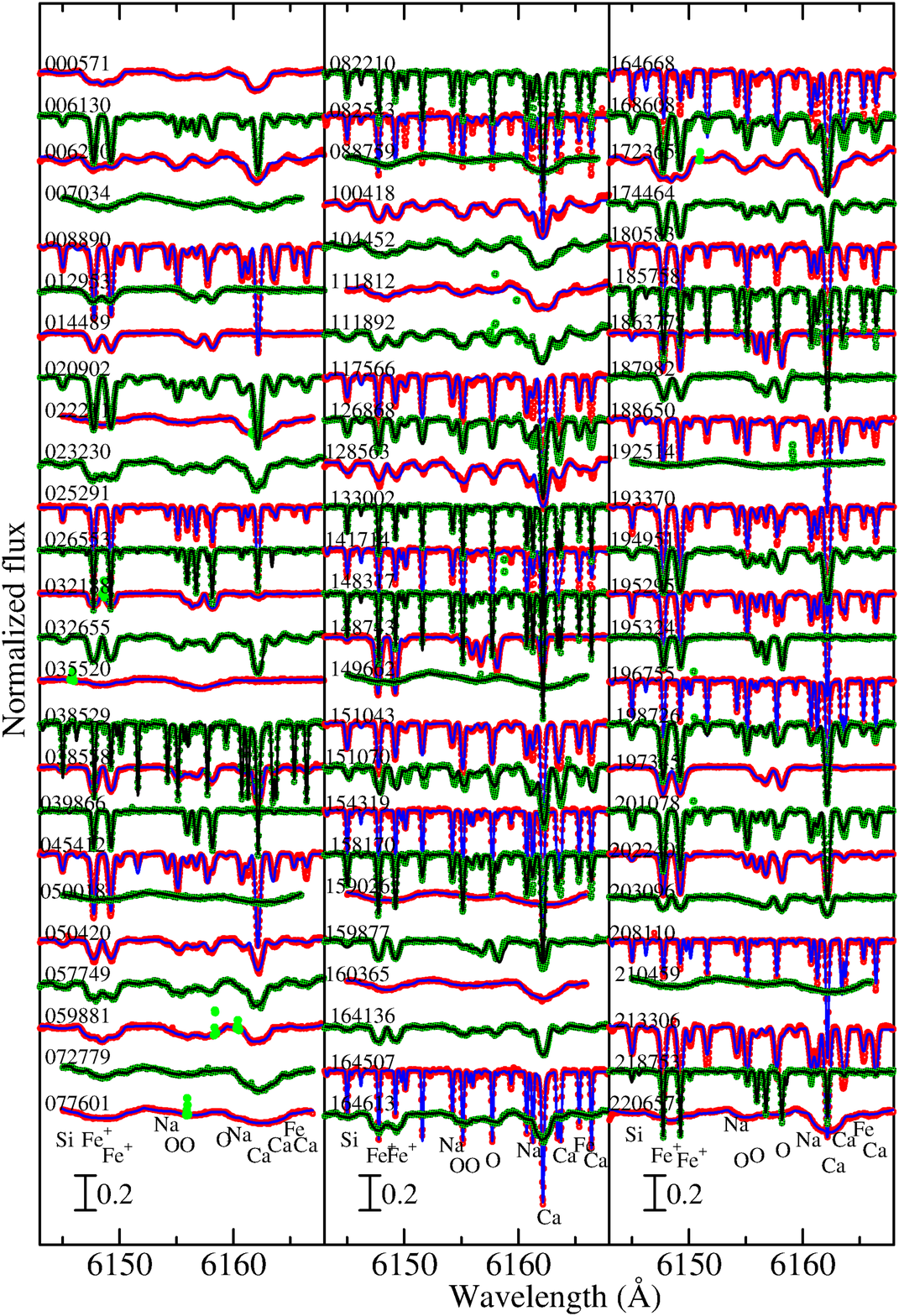}
\end{center}
\caption{
Synthetic spectrum fitting at the 6143--6168~$\rm\AA$ region.
Otherwise, the same as in figure~4.
}
\label{fig:5}
\end{figure}

\setcounter{figure}{5}
\begin{figure}
\begin{center}
  \FigureFile(100mm,120mm){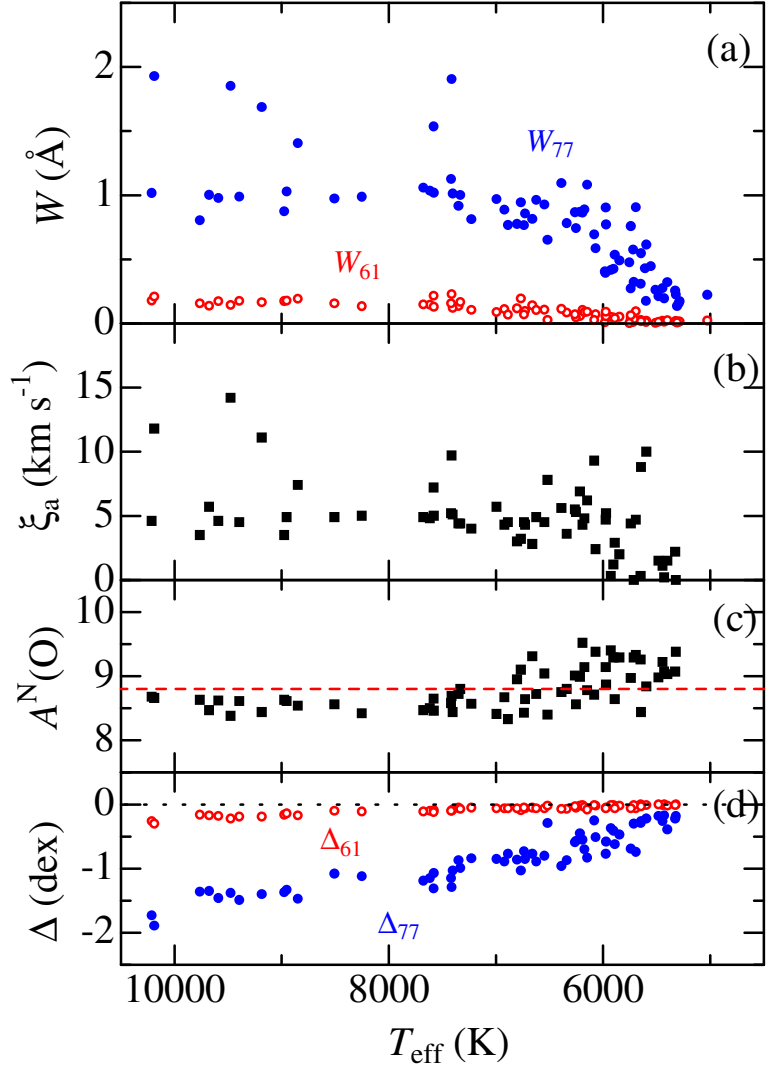}
\end{center}
\caption{
Oxygen abundances and related quantities for O~{\sc i} 7771--5 (suffix ``77'') 
and O~{\sc i} 6155--8 (suffix ``61'') lines, plotted against $T_{\rm eff}$. 
(a) Equivalent width $W_{77}$ (filled circles) and $W_{61}$ (open circles). 
(b) Abundance-based microturbulence $\xi_{\rm a}$, which was so determined that two 
non-LTE O abundances ($A^{\rm N}_{77}$ and $A^{\rm N}_{61}$) equal to each other.
(c) Non-LTE oxygen abundance $A^{\rm N}$(O) [= $A^{\rm N}_{77}$(O) = $A^{\rm N}_{61}$(O)] 
corresponding to $\xi_{\rm a}$.
(d) Non-LTE correction $\Delta_{77}$ (filled circles) and $\Delta_{61}$ (open circles),
}
\label{fig:6}
\end{figure}

\setcounter{figure}{6}
\begin{figure}
\begin{center}
  \FigureFile(120mm,120mm){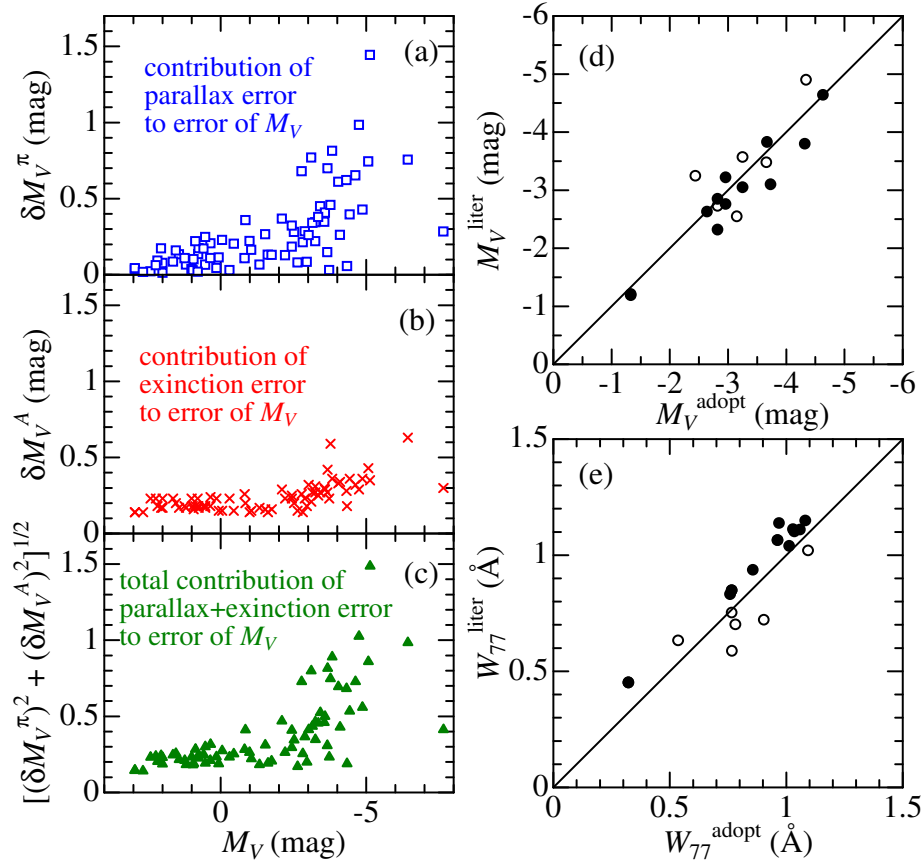}
\end{center}
\caption{
The left-hand panels show how the errors in parallax or interstellar extinction
contribute to errors in the absolute magnitude.
(a) Error component ($\delta M_{V}^{\pi}$) of $M_{V}$ due to error ($\sigma_{\pi}$) in 
parallax ($\pi$), which is defined as $\delta M_{V}^{\pi} \simeq 5\log (1 + \sigma_{\pi}/\pi)$, 
plotted against $M_{V}$.  
(b) Error component ($\delta M_{V}^{A}$) of $M_{V}$ due to error in 
interstellar extinction ($A_{V}$), plotted against $M_{V}$.
(c) Total error of $M_{V}$, which is defined as the root-sum-square of
$\delta M_{V}^{\pi}$ and $\delta M_{V}^{A}$, ploted against $M_{V}$.
In the right-hand panels are shown the comparison of our adopted values of $M_{V}$ (panel d) 
or $W_{77}$ (panel e) with those of Arellano Ferro et al. (2003) (6 stars in common: 
open symbols) and Kovtyukh et al. (2012) (11 stars in common: filed symbols)
}
\label{fig:7}
\end{figure}

\setcounter{figure}{7}
\begin{figure}
\begin{center}
  \FigureFile(120mm,140mm){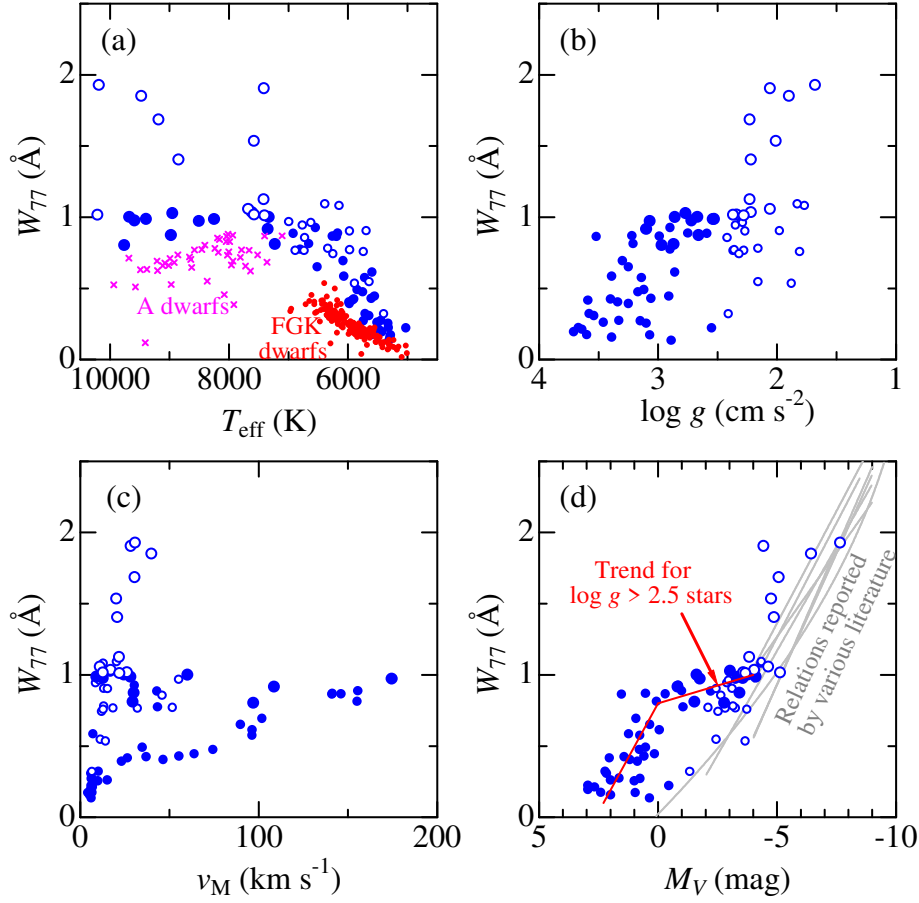}
\end{center}
\caption{
The strength of the whole O~{\sc i} 7771--5 multiplet ($W_{77}$)
plotted against (a) effective temperature ($T_{\rm eff}$), (b) surface gravity ($\log g$), 
(c) macrobroadening velocity ($v_{\rm M}$), and (d) absolute visual magnitude ($M_{V}$). 
Open and filled circles denote stars of lower $\log g (<2.5)$ and higher $\log g (> 2.5)$,
respectively, while lower $T_{\rm eff}$ ($< 7500$~K) and higher $T_{\rm eff}$ ($> 7500$~K) 
stars are expressed in smaller and larger symbols.
In panel (a), $W_{77}$ values of 160 FGK-type (red dots) and 46 A-type (pink crosses) 
main-sequence stars are shown for comparison.  In panel (d), gray solid lines represent 
the seven published $W_{77}$ vs. $M_{V}$ relations given in table 1 (Ref. 1, 2, 3, 7, 11, 12, 
and 13), while the red solid line is the eye-estimated mean trend for $\log g > 2.5$ stars 
(filled symbols).
}
\label{fig:8}
\end{figure}

\setcounter{figure}{8}
\begin{figure}
\begin{center}
  \FigureFile(120mm,160mm){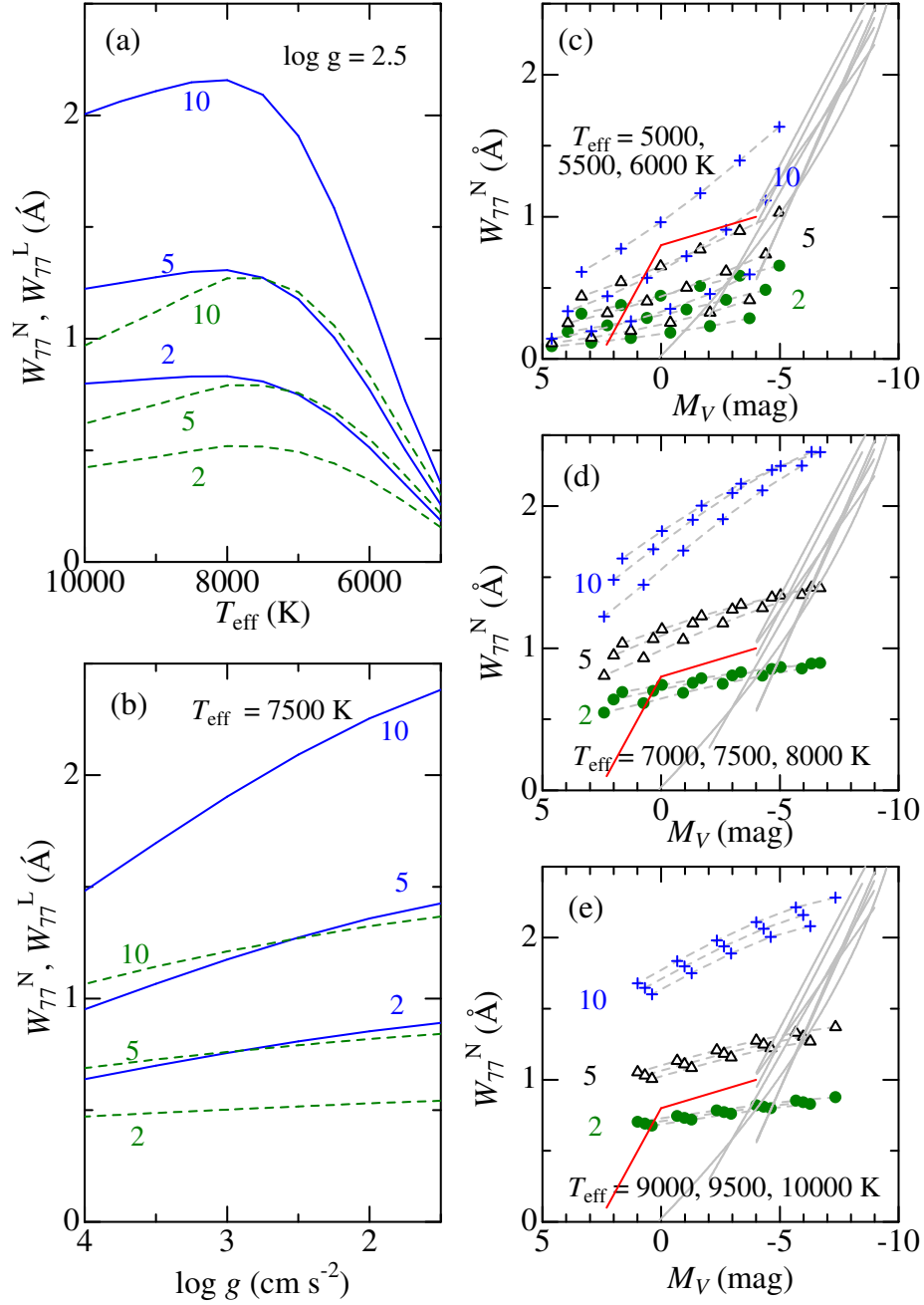}
\end{center}
\caption{
Graphical display of how the theoretical $W_{77}$ (computed for the solar oxygen abundance)
depends on the stellar parameters. 
Panels (a) and (b) show the $T_{\rm eff}$-dependence and $\log g$-dependence of $W_{77}$
for three microturbulence values of 2, 5, and 10~km~s$^{-1}$ (indicated in the figure), 
where the solid and dashed lines represent non-LTE ($W^{\rm N}_{77}$) and LTE ($W^{\rm L}_{77}$) 
results, respectively.
Panels (c), (d), and (e) illustrate the simulated dependence of $W^{\rm N}_{77}$ upon $M_{V}$ 
(evaluated by using equation~(1) and Alonso et al.'s (1999) bolometric correction) for 
three $T_{\rm eff}$ groups (5000--6000~K, 7000--8000~K, and 9000--10000~K; the data points
corresponding to the same $T_{\rm eff}$ but different $\log g$ are connected by gray dashed
lines), where the meanings of the solid lines are the same as in figure~8d. 
}
\label{fig:9}
\end{figure}

\setcounter{figure}{9}
\begin{figure}
\begin{center}
  \FigureFile(110mm,160mm){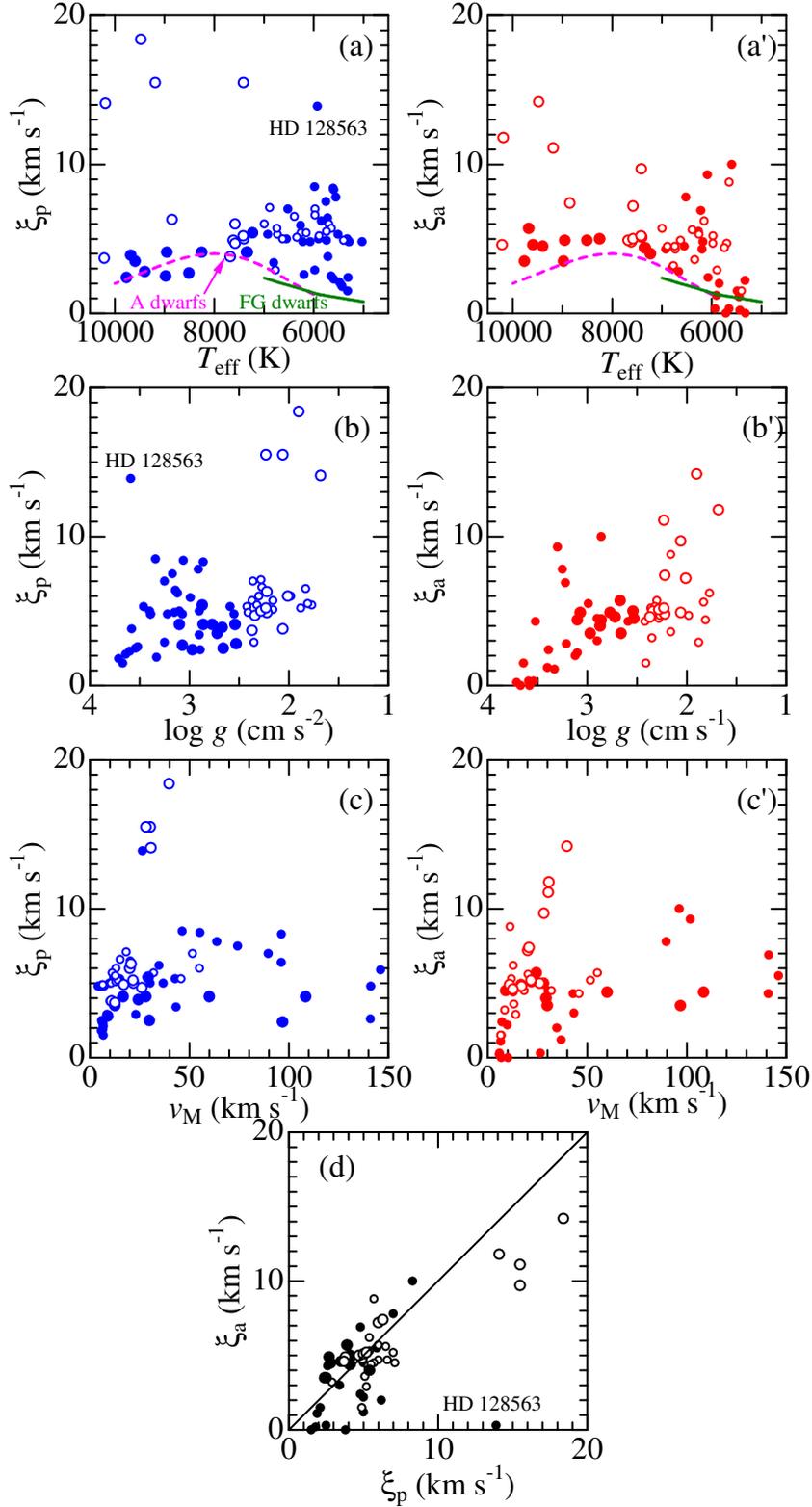}
\end{center}
\caption{
Microturbulence values plotted against (a) $T_{\rm eff}$, (b) $\log g$,
and (c) $v_{\rm M}$, where the left panels (a, b, c) are for $\xi_{\rm p}$
(profile-based microturbulence) while the right panels (a$'$, b$'$, c$'$)
are for $\xi_{\rm a}$ (abundance-based microturbulence). 
The solid and dashed line in panels (a) and (a$'$) are the mean trend of $\xi$ 
as a function of $T_{\rm eff}$ for F--G type main-sequence stars [Eq.~(1) and (2) 
of Takeda et al. (2013) for the case of $\log g = 4.0$] and A-type main-sequence 
stars [Eq.~(1) of Takeda et al. (2008a)], respectively. The correlation between
$\xi_{\rm p}$ and $\xi_{\rm a}$ is depicted in the bottom panel (d).
The outlier result for HD~128563 is unreliable, since sharp components  
are weakly overlapped at the line positions. 
See the caption of figure~8 for the meanings of the symbols.  
}
\label{fig:10}
\end{figure}

\setcounter{table}{0}
\begin{table}[h]
\footnotesize
\caption{Representative work related to O~{\sc i} 7771--5 line strength in A--F--G stars.}
\begin{center}
\begin{tabular}{ccccl}\hline\hline
No. & Reference & Star type$^{*}$ & Range of $M_{V}$ (mag) & $M_{V}$ (mag) vs. $W_{77}$ (\AA) relation \\
\hline
  (1) &  Osmer (1972)                   &F sg            &   from $-4$ to $-9$      &$M_{V} = -2.62W_{77} -2.55$ \\
  (2) &  Baker (1974)                   &F sg            &   from $-4$ to $-9$      &$M_{V} = -2.711W_{77} - 2.472$ \\
  (3) &  Sorvari (1974)                 &F sg            &   from $-2$ to $-9$      &$M_{V} = -3.42W_{77} -1.00$ \\
  (4) &  Kameswara Rao \& Mallik (1978)  &F--G sg          &   from $-2$ to $-10$     &$M_{V} = -10 E -1.79$ \\
  (5) &  Hopkinson \& Humrich (1981)     &A--G sg          &   N/A                &N/A \\
  (6) &  Faraggiana et al. (1988)       &B--F d--subg--g--sg &   N/A                &N/A \\
  (7) &  Arellano Ferro et al. (1989)    &F sg            &   from $-4$ to $-9$      &$M_{V} = -88.02\Lambda(16) + 187.83$ \\
  (8) &  Arellano Ferro et al. (1991)    &F--G g--sg        &   from $+2$ to $-10$     &$M_{V} = 1.52 -6.33 W_{77} + 0.85 W_{77}^{2} - 3.74 (b-y)_{0}$ \\
  (9) &  Arellano Ferro \& Mendoza (1993)&A--G g--sg        &   from $+3$ to $-10$     &$M_{V} = 8.0 -13.3 W_{77} + 2.7 W_{77}^{2} - 2.1 (b-y)_{0}$ \\
  (10)&  Mendoza \& Arellano Ferro (1993)&A--G g--sg        &   from $+3$ to $-8$      &N/A \\
  (11)&  Slowik \& Peterson (1993)       &A--F sg          &   from $-4$ to $-8.5$    &$M_{V}=-1.07-3.11W_{77}$ \\
  (12)&  Slowik \& Peterson (1995)       &A-F sg          &   from $-4$ to $-9$      &$M_{V}=-0.68-3.17W_{77}$ \\
  (13)&  Arellano Ferro et al. (2003)   &A--G g--sg, Cep  &   from $+0.35$ to $-9.5$ &$M_{V} = 0.131 -5.831 W_{77} + 0.789 W_{77}^{2}$ \\
  (14)&  Kovtyukh et al. (2012)         &F--G sg          &   from $0$ to $-9$       & [cf. their Eq.(2)] \\
  (15)&  Dambis (2013)                  &F sg            &   from $M_{K}$  = $-4.7$ to $-9$ &$M_{K}=-5.33-10.81\log W_{77}$ \\
\hline
\end{tabular}
\end{center}
$^{*}$Meaning of abbreviation: `d' $\cdots$ dwarfs, `subg' $\cdots$ subgiants, `g' $\cdots$ giants, `sg' $\cdots$ supergiants, and `Cep' $\cdots$ Cepheids.\\
Observational method and specific remark.\\
(1) Photoelectric scanner. No clear correlation was found for A supergiants.\\
(2) Photoelectric scanner.\\
(3) Narrow-band photometry.\\
(4) Photographic spectroscopy (16 or 33 \AA~mm$^{-1}$). $E \equiv W$(O~{\sc i}~7771--5)$\times W$(Fe~{\sc i}~7748).\\
(5) Spectroscopy with linear photodiode array (10 \AA~mm$^{-1}$).\\
(6) Spectroscopy with Reticon (50 \AA~mm$^{-1}$). \\
(7) Narrow-band photometry. $W_{77} = 20.8 \Lambda (16)-44.3$.\\
(8) High-resolution CCD spectroscopy (8 \AA~mm$^{-1}$).\\
(9) Narrow-band photometry and low-resolution spectroscopy.\\
(10) Narrow-band photometry and CCD spectroscopy. Application of Arellano Ferro \& Mendoza (1993) formula to derive $M_{V}$.\\
(11) Medium-resolution CCD spectroscopy.\\
(12) Medium-resolution CCD spectroscopy.\\
(13) High-resolution CCD spectroscopy ($R=18000$).\\
(14) High-dispersion spectroscopy ($R=52000$ or 85000). Their Eq.(2) is an intricate relation in terms of $T_{\rm eff}$, $\log g$, $\xi$, and [Fe/H].\\
(15) Use of published $W_{77}$ data from various literature.\\

\end{table}

\setcounter{table}{1}
\footnotesize
\renewcommand{\arraystretch}{0.8}
\setlength{\tabcolsep}{3pt}
\begin{longtable}{ccccccccccccl}
\caption{Basic stellar parameters and the results of equivalent widths 
as well as microturbulences for 75 program stars.}
\hline\hline
HD\# &  Name & Sp.Type & $M_{V}$ & $\log L$ & $M$ & $T_{\rm eff}$ & 
$\log g$ & $v_{\rm M}$ & $\xi_{\rm p}$ & $W_{77}$ & $\xi_{\rm a}$ &Remark \\
(1) & (2) & (3) & (4) & (5) & (6) & (7) & (8) & (9) & (10) & (11) & (12) & (13)\\
\hline
\endhead
\hline
\endfoot
\hline
\multicolumn{13}{l}{\hbox to 0pt{\parbox{150mm}{\footnotesize
(1) HD number. (2) Bayer/Flamsteed name. (3) Spectral type taken from Hipparcos catalogue (ESA 1997).
(4) Extinction-corrected absolute visual magnitude (in mag). (5) Logarithmic bolometric
luminosity $\log (L/L_{\odot})$ (in dex), where $L_{\odot}$ is the solar luminosity.
(6) Stellar mass (in $M_{\odot}$). 
(7) Effective temperature (in K). (8) Logarithm of surface gravity $\log g$ (in dex),
where $g$ is in unit of cm~s$^{-2}$. (9) Macrobroadening velocity in km~s$^{-1}$ (nearly 
equivalent to the projected rotational velocity $v_{\rm e}\sin i$ for large $v_{\rm M}$, 
though the contribution of macroturbulence may be significant for the case of small $v_{\rm M}$ 
with $\ltsim$ several tens~km~s$^{-1}$). (10) Profile-based microturbulence (in km~s$^{-1}$). 
(11) Equivalent width of the whole O~{\sc i} 7771--5 triplet
(in $\rm\AA$). (12) Abundance-based microturbulence (in km~s$^{-1}$).
(13) Specific remark. The notation 
``1st Hipp. plx''means that the parallax data was taken from the first version of the 
Hipparcos catalogue (ESA 1997), instead of the revised new reduction data (van Leeuwen 2007).
}}}
\endlastfoot
\hline
000571 &22 And           &F2~II         & $-$3.25 & 3.202&  5.42 &  6995 & 2.30 &  55 &   6.0 & 0.969 &   5.7 & \\
006130 &                 &F0~II         & $-$4.04 & 3.519&  6.54 &  7616 & 2.22 &  17 &   4.9 & 1.035 &   4.8 & \\
006210 &                 &F6~V          & +1.19 & 1.451&  1.97 &  5983 & 3.34 &  46 &   8.5 & 0.405 &  $\cdots$ & \\
007034 &82 Psc           &F0~V          & $-$0.82 & 2.229&  2.97 &  7349 & 3.10 & 109 &   4.1 & 0.917 &   4.4 & \\
008890 &$\alpha$ UMi     &F7:~Ib-IIv SB & $-$3.73 & 3.428&  6.50 &  5741 & 1.81 &  13 &   5.5 & 0.759 &   4.4 &Cepheid \\
012953 &                 &A1~Ia         & $-$6.43 & 4.502& 12.83 &  9475 & 1.90 &  40 &  18.4 & 1.852 &  14.2 & \\
014489 &9 Per            &A2~Ia         & $-$5.07 & 3.949&  8.54 &  9185 & 2.23 &  30 &  15.5 & 1.686 &  11.1 &1st Hipp. plx \\
020902 &$\alpha$ Per     &F5~Ib         & $-$4.34 & 3.650&  7.35 &  6389 & 1.83 &  20 &   6.5 & 1.094 &   5.6 & \\
022211 &                 &G0           & +0.77 & 1.633&  2.27 &  5720 & 3.14 &  96 &   6.4 & 0.576 &  $\cdots$ & \\
023230 &41 Per           &F5~IIvar      & $-$2.64 & 2.960&  4.71 &  6728 & 2.42 &  46 &   5.3 & 0.857 &   4.3 &double line? \\
025291 &                 &F0~II         & $-$4.63 & 3.755&  7.65 &  7677 & 2.06 &  11 &   3.8 & 1.059 &   4.9 &1st Hipp. plx \\
026553 &                 &A4~III        & $-$2.89 & 3.059&  5.00 &  6767 & 2.35 &   9 &   2.9 & 0.945 &   3.2 & \\
032188 &                 &A2sh         & $-$3.42 & 3.287&  5.47 &  8976 & 2.66 &  30 &   2.5 & 0.875 &   3.5 & \\
032655 &                 &F2~IIp...     & $-$0.85 & 2.248&  3.08 &  6548 & 2.90 &  30 &   5.0 & 0.927 &   4.5 & \\
035520 &                 &A1p          & $-$2.78 & 3.049&  4.69 &  9763 & 2.97 &  97 &   2.4 & 0.804 &   3.5 &1st Hipp. plx \\
038529 &                 &G4~V          & +2.95 & 0.790&  1.46 &  5320 & 3.67 &   7 &   1.5 & 0.225 &   0.0 & \\
038558 &130 Tau          &F0~III        & $-$3.58 & 3.333&  5.80 &  7579 & 2.34 &  26 &   4.7 & 1.019 &   5.0 & \\
039866 &                 &A2~II         & $-$5.13 & 4.000&  8.73 & 10212 & 2.37 &  13 &   3.7 & 1.017 &   4.6 &1st Hipp. plx \\
045412 &48 Aur           &F5.5~Ibv      & $-$3.11 & 3.186&  5.60 &  5694 & 1.98 &  13 &   6.0 & 0.905 &   4.7 &Cepheid, 1st Hipp. plx \\
050018 &59 Aur           &F2~V          & +0.07 & 1.879&  2.51 &  6660 & 3.21 & 155 &  $\cdots$ & 0.814 &   2.8 & \\
050420 &                 &A9~III        & $-$1.53 & 2.511&  3.56 &  7229 & 2.87 &  29 &   5.4 & 0.812 &   4.0 & \\
057749 &                 &F3~IV         & $-$1.05 & 2.325&  3.21 &  6803 & 2.90 &  43 &   3.4 & 0.775 &   3.0 & \\
059881 &$\delta^{1}$ CMi &F0~III        & $-$1.62 & 2.550&  3.63 &  7332 & 2.86 &  60 &   4.1 & 1.002 &   4.4 & \\
072779 &35 Cnc           &G0~III        & $-$0.05 & 1.967&  2.78 &  5596 & 2.86 &  96 &   8.3 & 0.614 &  10.0 & \\
077601 &                 &F6~II-III     & +0.53 & 1.707&  2.30 &  6216 & 3.22 & 141 &   4.8 & 0.870 &   6.9 & \\
082210 &24 UMa           &G4~III-IV     & +2.00 & 1.173&  1.88 &  5294 & 3.39 &   5 &   4.8 & 0.157 &  $\cdots$ & \\
082543 &                 &F7~IV-V       & +0.76 & 1.633&  2.26 &  5742 & 3.15 &   6 &   4.9 & 0.273 &  $\cdots$ & \\
088759 &                 &F2           & +0.35 & 1.770&  2.36 &  6518 & 3.25 &  90 &   7.0 & 0.652 &   7.8 & \\
100418 &                 &F8/G0~Ib/II   & +0.53 & 1.721&  2.39 &  5847 & 3.12 &  35 &   6.2 & 0.492 &   2.0 & \\
104452 &1 Com            &G0~II         & +0.59 & 1.710&  2.41 &  5608 & 3.06 &  55 &   8.4 & 0.430 &  $\cdots$ & \\
111812 &31 Com           &G0~III        & +0.15 & 1.889&  2.69 &  5554 & 2.91 &  64 &   7.8 & 0.446 &  $\cdots$ & \\
111892 &33 Com           &F8           & +1.43 & 1.357&  1.89 &  5903 & 3.40 &  37 &   5.0 & 0.426 &   1.2 & \\
117566 &                 &G2.5~IIIb     & +1.00 & 1.569&  2.34 &  5327 & 3.10 &  10 &   5.0 & 0.256 &   2.2 & \\
126868 &$\phi$ Vir       &G2~III        & +2.00 & 1.154&  1.79 &  5511 & 3.46 &  15 &   5.3 & 0.262 &  $\cdots$ & \\
128563 &                 &F8           & +2.05 & 1.109&  1.63 &  5927 & 3.59 &  26 &  13.9 & 0.417 &   0.3 &sharp component \\
133002 &                 &F9~V          & +2.41 & 0.985&  1.58 &  5599 & 3.60 &   7 &   2.3 & 0.175 &  $\cdots$ & \\
141714 &$\delta$ CrB     &G5~III-IV     & +0.96 & 1.590&  2.38 &  5284 & 3.07 &   4 &   4.8 & 0.173 &  $\cdots$ & \\
148317 &                 &G0~III        & +2.17 & 1.076&  1.66 &  5649 & 3.54 &   6 &   2.5 & 0.309 &   0.3 & \\
148743 &                 &A7~Ib         & $-$4.75 & 3.800&  7.91 &  7581 & 2.01 &  20 &   6.0 & 1.536 &   7.2 & \\
149662 &                 &F2           & +1.54 & 1.300&  1.80 &  6192 & 3.52 & 141 &   2.6 & 0.865 &   4.3 & \\
151043 &                 &F8           & +2.23 & 1.049&  1.61 &  5713 & 3.58 &  10 &   3.8 & 0.323 &   0.0 & \\
151070 &                 &F5~III        & +0.87 & 1.578&  2.15 &  5977 & 3.25 &  23 &   2.9 & 0.394 &  $\cdots$ &double line? \\
154319 &                 &K0           & +1.64 & 1.301&  1.98 &  5445 & 3.33 &   7 &   1.9 & 0.275 &   1.1 & \\
158170 &                 &F5~IV         & +1.24 & 1.426&  1.94 &  6069 & 3.39 &   7 &   4.8 & 0.586 &   2.4 & \\
159026 &                 &F6~III        & $-$0.99 & 2.313&  3.25 &  6174 & 2.75 & 156 &  $\cdots$ & 0.889 &   4.8 & \\
159877 &                 &F0~IV         & $-$3.77 & 3.441&  5.98 &  9676 & 2.67 &  24 &   3.9 & 1.002 &   5.7 & \\
160365 &                 &F6~III        & +0.93 & 1.550&  2.09 &  6083 & 3.30 & 102 &  $\cdots$ & 0.695 &   9.3 & \\
164136 &$\nu$ Her        &F2~II         & $-$2.82 & 3.032&  4.92 &  6738 & 2.37 &  32 &   5.7 & 0.766 &   4.5 & \\
164507 &                 &G5~IV         & +2.95 & 0.780&  1.43 &  5429 & 3.71 &   6 &   1.8 & 0.196 &   0.2 & \\
164613 &$\psi^{2}$ Dra   &F2.5~II-III   & $-$2.21 & 2.785&  4.21 &  6921 & 2.59 &  43 &   5.3 & 0.887 &   4.3 & \\
164668 &95 Her B         &G5           & $-$0.45 & 2.179&  3.37 &  5027 & 2.55 &   6 &   4.8 & 0.223 &  $\cdots$ &$V$ and $B-V$ from simbad \\
168608 &Y Sgr            &F8~II         & $-$3.25 & 3.201&  5.43 &  6887 & 2.28 &  18 &   7.1 & 0.767 &   4.5 &Cepheid \\
172365 &                 &F8~Ib-II      & $-$2.10 & 2.768&  4.32 &  5972 & 2.36 &  52 &   7.0 & 0.771 &   5.2 & \\
174464 &                 &F2~Ib         & $-$3.67 & 3.368&  5.96 &  7403 & 2.28 &  22 &   5.0 & 1.013 &   5.1 &1st Hipp. plx \\
180583 &V473 Lyr         &F6~Ib-II      & $-$2.52 & 2.925&  4.69 &  6253 & 2.32 &  12 &   5.4 & 0.743 &   5.3 &Cepheid \\
185758 &$\alpha$ Sge     &G0~II         & $-$1.33 & 2.494&  3.83 &  5400 & 2.41 &   7 &   4.9 & 0.322 &   1.5 & \\
186377 &                 &A5~III        & $-$3.56 & 3.357&  5.67 &  9590 & 2.72 &  13 &   3.5 & 0.978 &   4.6 & \\
187982 &                 &A1~Iab        & $-$4.43 & 3.674&  7.28 &  7413 & 2.06 &  28 &  15.5 & 1.906 &   9.7 & \\
188650 &                 &Fp           & $-$2.44 & 2.921&  4.80 &  5645 & 2.16 &  11 &   5.7 & 0.547 &   8.8 & \\
192514 &30 Cyg           &A5~IIIn       & $-$1.75 & 2.610&  3.67 &  8507 & 3.07 & 175 &   2.7 & 0.974 &   4.9 & \\
193370 &35 Cyg           &F5~Ib         & $-$4.32 & 3.648&  7.39 &  6149 & 1.77 &  13 &   5.4 & 1.082 &   6.2 &1st Hipp. plx \\
194951 &                 &F1~II         & $-$3.83 & 3.434&  6.21 &  7418 & 2.23 &  22 &   5.2 & 1.126 &   5.2 &1st Hipp. plx \\
195295 &41 Cyg           &F5~II         & $-$2.96 & 3.092&  5.13 &  6624 & 2.29 &  11 &   5.0 & 0.963 &   4.9 & \\
195324 &42 Cyg           &A1~Ib         & $-$4.87 & 3.862&  8.06 &  8849 & 2.22 &  21 &   6.3 & 1.405 &   7.4 & \\
196755 &$\kappa$ Del     &G5~IV+...     & +2.67 & 0.889&  1.51 &  5482 & 3.64 &   7 &   2.1 & 0.213 &   1.5 & \\
197345 &$\alpha$ Cyg     &A2~Ia         & $-$7.65 & 5.006& 18.33 & 10188 & 1.68 &  31 &  14.1 & 1.928 &  11.8 & \\
198726 &T Vul            &F5~Ib         & $-$2.44 & 2.902&  4.68 &  5974 & 2.27 &  15 &   6.6 & 0.904 &   4.7 &Cepheid \\
201078 &DT Cyg           &F7.5~Ib-IIv   & $-$3.15 & 3.173&  5.44 &  6339 & 2.16 &  13 &   5.1 & 0.782 &   3.6 &Cepheid \\
202240 &                 &F0~III        & $-$3.02 & 3.128&  4.96 &  8952 & 2.77 &  17 &   4.1 & 1.029 &   4.9 & \\
203096 &                 &A5~IV         & $-$3.35 & 3.247&  5.41 &  8251 & 2.54 &  28 &   4.1 & 0.987 &   5.0 & \\
208110 &                 &G0~IIIs       & +0.36 & 1.828&  2.70 &  5309 & 2.89 &   6 &   2.4 & 0.136 &  $\cdots$ & \\
210459 &$\pi$ Peg        &F5~III        & $-$0.30 & 2.035&  2.78 &  6262 & 2.99 & 146 &   5.9 & 0.867 &   5.5 & \\
213306 &$\delta$ Cep     &G2~Ibvar      & $-$3.66 & 3.393&  6.32 &  5890 & 1.88 &  14 &   5.2 & 0.536 &   2.9 &Cepheid \\
218753 &2 Cas            &A5~III        & $-$4.10 & 3.567&  6.52 &  9395 & 2.53 &   9 &   2.8 & 0.988 &   4.5 & \\
220657 &$\upsilon$ Peg   &F8~IV         & +0.79 & 1.621&  2.28 &  5754 & 3.17 &  74 &   7.5 & 0.476 &  $\cdots$ & \\
\end{longtable}

\setcounter{table}{2}
\begin{table}[h]
\footnotesize
\caption{Adopted atomic data of oxygen lines.}
\begin{center}
\begin{tabular}{cccccccc}\hline\hline
Line & Equivalent & $\lambda$ & $\chi_{\rm low}$ & $\log gf$ & Gammar & Gammas & Gammaw\\
     &  Width  & ($\rm\AA$) & (eV) & (dex) & (dex) & (dex) & (dex) \\  
\hline
O~{\sc i} 7771--5 & $W_{77}$ & 7771.944 &  9.146 & $+0.324$ & 7.52 & $-$5.55 & ($-$7.65)\\
            & (3 components) & 7774.166 &  9.146 & $+0.174$ & 7.52 & $-$5.55 & ($-$7.65)\\
                  &          & 7775.388 &  9.146 & $-0.046$ & 7.52 & $-$5.55 & ($-$7.65)\\
\hline
O~{\sc i} 6155--8 & $W_{61}$ & 6155.961 & 10.740 & $-1.401$ & 7.60 & $-$3.96 & ($-$7.23)\\
            & (9 components) & 6155.971 & 10.740 & $-1.051$ & 7.61 & $-$3.96 & ($-$7.23)\\
                  &          & 6155.989 & 10.740 & $-1.161$ & 7.61 & $-$3.96 & ($-$7.23)\\
                  &          & 6156.737 & 10.740 & $-1.521$ & 7.61 & $-$3.96 & ($-$7.23)\\
                  &          & 6156.755 & 10.740 & $-0.931$ & 7.61 & $-$3.96 & ($-$7.23)\\
                  &          & 6156.778 & 10.740 & $-0.731$ & 7.62 & $-$3.96 & ($-$7.23)\\
                  &          & 6158.149 & 10.741 & $-1.891$ & 7.62 & $-$3.96 & ($-$7.23)\\
                  &          & 6158.172 & 10.741 & $-1.031$ & 7.62 & $-$3.96 & ($-$7.23)\\
                  &          & 6158.187 & 10.741 & $-0.441$ & 7.61 & $-$3.96 & ($-$7.23)\\ 
\hline
\end{tabular}
\end{center}
Following columns 3--5 (laboratory wavelength, lower excitation potential, 
and statistical weight of lower level times oscillator strength), 
damping parameters are presented in columns 6--8:  
Gammar is the radiation damping width (s$^{-1}$) [$\log\gamma_{\rm rad}$], 
Gammas is the Stark damping width (s$^{-1}$) per electron density (cm$^{-3}$) 
at $10^{4}$ K [$\log(\gamma_{\rm e}/N_{\rm e})$], and
Gammaw is the van der Waals damping width (s$^{-1}$) per hydrogen density 
(cm$^{-3}$) at $10^{4}$ K [$\log(\gamma_{\rm w}/N_{\rm H})$]. \\
These data were taken from the compilation of Kurucz and Bell (1995)
as long as available, while the parenthesized damping parameters are 
the default values computed by the WIDTH9 program.
\end{table}

\end{document}